\documentstyle[colordvi,psfig]{mn}   
\def\om{\Omega_p}
\def\len{a_B}
\def\lag{D_L}
\def\vpd{{\cal R}}
\def\kin{{\cal V}}
\def\pin{{\cal X}}
   
\def\hi{H~{\small I}}   
\def\gtsim{\mathrel{\spose{\lower.5ex \hbox{$\mathchar"218$}}
     \raise.4ex\hbox{$\mathchar"13E$}}}
\def\ltsim{\mathrel{\spose{\lower.5ex\hbox{$\mathchar"218$}}
     \raise.4ex\hbox{$\mathchar"13C$}}}

\def\degrees{^\circ}
\def\arcmin{^\prime}
\def\arcsec{^{\prime\prime}}
\def\kms{$\mathrm km\;s^{-1}$}
\def\kmsa{$\mathrm km\;s^{-1}\,arcsec^{-1}$}
\def\kmsm{$\mathrm km\;s^{-1}\,Mpc^{-1}$}
\def\kmsk{$\mathrm km\;s^{-1}\,kpc^{-1}$}
\def\msun{M$_\odot$}
\def\mas{mag arcsec$^{-2}$}   

\def\iraf{{\tt IRAF}}
\def\midas{{\tt MIDAS}}
\def\etal{{et al.}}
\def\eg{{\it e.g.}}

\def\ie{{\it i.e.}}
\def\cf{{\it cf.}}

\def\spose#1{\hbox to 0pt{#1\hss}}

\begin{document}   
   
\title[A Fast Bar in the Post-Interaction Galaxy NGC 1023]{A Fast Bar 
in the Post-Interaction Galaxy NGC 1023 \thanks{Based on observations 
made with the Italian Telescopio
    Nazionale Galileo (AOT-3, 3-06-119) operated on the island of La
    Palma by the Centro Galileo Galilei of the Consorzio Nazionale per
    l'Astronomia e l'Astrofisica, and the Jacobus Kapteyn Telescope
    operated by the Isaac Newton group at La Palma island  at the Spanish 
    Observatorio del Roque
    de los Muchachos of the Instituto de Astrofisica de Canarias.}}

   
\author[V.P. Debattista, E.M. Corsini and J.A.L. Aguerri] 
{Victor P. Debattista$^1$\thanks{email: {\tt debattis@astro.unibas.ch}}, 
Enrico Maria Corsini$^2$,
and J. A. L. Aguerri$^1$\\
$^1$ Astronomisches Institut, Universit\"at Basel, Venusstrasse 7,
CH-4102 Binningen, Switzerland\\ 
$^2$ Osservatorio Astrofisico di Asiago, Dipartimento di Astronomia,
Universit\`a di Padova, via dell'Osservatorio~8, I-36012 Asiago, Italy\\
}

\date{{\it Draft version on \today}}

\maketitle   
   
\begin{abstract}   

We measured the bar pattern speed, $\om$, of the SB0 galaxy NGC 1023
using the Tremaine-Weinberg (1984) method with stellar-absorption slit 
spectroscopy.  The morphology and kinematics 
of the \hi\ gas outside NGC 1023 suggest it suffered a tidal interaction, 
sometime in the past, with one of its dwarf companions.  At present, however, 
the optical disc is relaxed.  If the disc had been stabilized by a massive 
dark matter halo and formed its bar in the interaction, then the bar would 
have to be slow.  We found $\om = 5.0 \pm 1.8$ \kmsa, so that the bar ends 
near its co-rotation radius.  It is therefore rotating rapidly and must have a 
maximum disc.

\end{abstract}   
   

\section{Introduction}   
\label{sec:introduction}   
   
Strong bars are seen in optical images of roughly 30\% of all high surface
brightness (HSB) disk galaxies (Sellwood \& Wilkinson 1993) and this fraction 
rises to 50\%-75\% in the near IR (Knapen 1999; Knapen \etal\ 2000; Eskridge 
\etal\ 2000).  Understanding the structure and dynamics of barred (SB) 
galaxies is, therefore, an issue of some importance.  The principal dynamical 
quantity for SB galaxies is the rotation frequency/pattern speed of the bar, 
$\om$. This is usually parametrized by the distance-independent ratio 
$\vpd \equiv \lag/\len$, where $\len$ is the semi-major axis of the bar and 
$\lag$ is the radius to the Lagrangian point, where the gravitational and 
centrifugal forces cancel out in the bar's rest frame.  (The Lagrangian 
radius is therefore the generalization to strong bars of the corotation 
radius.)  Contopoulos (1980) argued that a self-consistent bar is possible 
only if $\vpd \geq 1$.  A bar is termed fast when $1.0 \leq \vpd \ltsim 1.4$, 
while, for a larger value of $\vpd$, a bar is said to be slow.

A variety of methods have been used to attempt measurent of bar pattern 
speeds (see, for example, the review of Elmegreen 1996).  Most measurements 
of $\vpd$ rely on hydrodynamical simulations.  These
usually try to match the gas flow in the region of the bar,
particularly at the shocks, which works because the location of the
shocks depends on $\vpd$, moving further ahead of the bar as $\vpd$
increases.  A bar needs to be fast for the shocks to remain curved
with their concave sides towards the bar major axis, as is usually
observed (van Albada \& Sanders 1982; Athanassoula 1992).  Detailed 
simulations of gas flows
in individual galaxies also result in fast bars; examples include: NGC
1365 ($\vpd = 1.3$, Lindblad \etal\ 1996), NGC 1300 ($\vpd = 1.3$,
Lindblad \& Kristen 1996), and NGC 4123 ($\vpd = 1.2$, Weiner \etal\ 
2001).  Hydrodynamical simulations can also recover $\vpd$ by matching
morphological features in \hi; some examples are: NGC 7479 ($\vpd
=1.22$, Laine 1996), NGC 1073 ($\vpd =1 - 1.2$, England \etal\ 1990),
NGC 3992 ($\vpd =1$, Hunter \etal\ 1989), and NGC 5850 $(\vpd =1.35$,
Aguerri \etal\ 2001).

A direct method for measuring $\om$, using a tracer population which
satisfies continuity, was developed by Tremaine \& Weinberg (1984).
Since gas is subject to phase changes, it is not well-suited for this
application.  Old stellar populations in the absence of significant
obscuration, on the other hand, are ideal for the Tremaine-Weinberg
(TW) method.  This has permitted application of the method to a small
number of early type SB galaxies: NGC 936 ($\vpd = 1.4 \pm 0.3$, Kent
1987 and Merrifield \& Kuijken 1995), NGC 4596 ($\vpd =
1.15^{+0.38}_{-0.23}$, Gerssen \etal \ 1998) and NGC 7079 ($\vpd = 0.9
\pm 0.15$, Debattista \& Williams 2001).

The observational evidence, therefore, favors fast bars.  The
perturbation theory calculations of Weinberg (1985) predicted that a
fast bar would be slowed down rapidly in the presence of a massive
dark matter (DM) halo.  Such slow-down has been seen in various
simulations (Sellwood 1980; Little \& Carlberg 1991; Hernquist \&
Weinberg 1992; Athanassoula 1996).  The fully self-consistent, high
resolution $N$-body simulations of Debattista \& Sellwood (1998) also
confirmed this prediction; however they showed that, for a maximum
disc (here taken to mean a disc which dominates the rotation curve
throughout the inner few disc scale-lengths, \cf\ van Albada \&
Sancisi 1986), a fast bar can survive for a large fraction of a Hubble
time.  Subsequently, Tremaine \& Ostriker (1999) suggested that bars
manage to remain fast not because discs are maximal but rather because
the inner parts of DM halos are flattened and rapidly rotating.  However,
Debattista \& Sellwood (2000) showed that rapid slow-down persists even
then unless the halo angular momentum is very large relative to that of the
disc.  Thus they concluded that SB galaxies must be maximal, and
argued that the same must be true for all high surface brightness
disc galaxies.

This conclusion rests on a small number of pattern speed measurements;
in view of the fact that maximum discs are in conflict with the
predictions of cold dark matter (CDM) cosmologies (\eg\ Navarro \etal\ 
1997), enlarging the sample of measured pattern speeds is desireable.
In this paper, we report observations of NGC 1023, for which we
applied the TW method.

The rest of this paper is organized as follows. The TW method is
described briefly in \S\ref{sec:twmethod}.  Then, in
\S\ref{sec:properties} we give an overview of the previously known
properties of NGC 1023.  The photometric observations, reduction and
results, including $\len$, are presented in \S\ref{sec:photometry},
while \S\ref{sec:spectroscopy} presents the spectroscopic observations
and results.  We derive the rotation curve, corrected for the
asymmetric drift, from which we obtain $\lag$.  With these results at
hand, we then apply the TW method in \S\ref{sec:pattern_speed}.  We
present our conclusions in \S\ref{sec:conclusions}.

\section{The Tremaine-Weinberg Method}
\label{sec:twmethod}

The TW method is contained in the following simple equation:
\begin{equation}  
\om \sin i \ =  \ \frac{\int^\infty_{-\infty}~h(Y)
\int^\infty_{-\infty}\Sigma(X,Y)~V_{\it los}(X,Y)~dX~dY}
{\int^\infty_{-\infty}~h(Y)
\int^\infty_{-\infty}\Sigma(X,Y)~X~dX~dY}
\label{eq:TW_equation}
\end{equation}
where $i$ is the inclination of the galaxy, $V_{\it los}$ is the
line-of-sight (minus systemic) velocity, $\Sigma$ is the surface
brightness, $h$ is an arbitrary weighting function and $(X,Y)$ are
coordinates on the sky, centered on the galaxy, along the disc's
apparent major and minor axes, respectively.  Slit observations are
equivalent to $h(Y) \propto \delta(Y-Y_{\it slit})$.  Application of
Eqn.  \ref{eq:TW_equation} presents three main difficulties:

\begin{enumerate}
\item{\it Centering errors.}  As was already recognized by Tremaine \&
  Weinberg (1984), small errors in identifying either the center or
  the systemic velocity of the galaxy can significantly affect the
  value of $\om$ obtained.  Noting that most SB galaxies are nearly
  point-symmetric about their centers, Tremaine \& Weinberg (1984)
  suggested using a weighting function which is odd in $Y$ to counter
  this problem.  Merrifield \& Kuijken (1995) used a different, and
  somewhat better, strategy.  Noting that the integrals in Eqn.
  \ref{eq:TW_equation} represent luminosity-weighted averages, they
  rewrote Eqn. \ref{eq:TW_equation} as:
  \begin{equation}
  \om \sin i (\pin - X_c) = \kin - V_{sys},
  \end{equation}
  where $\kin$ and $\pin$ are the luminosity-weighted averages
  relative to an arbitrary frame, in which the galaxy center is
  $(X_c,Y_c)$ and the systemic velocity is $V_{sys}$.  Plotting $\kin$
  against $\pin$ then gives a straight line with slope $\om \sin i$.
  With this approach, the problem of centering errors becomes one of
  fixing an arbitrary reference position and velocity frame common to
  all the slits, which in general is much easier to achieve.
  
\item{\it Small signal-to-noise ratios.}  $\kin$ and $\pin$ measure
  differences across $X=0$ and are, therefore, susceptible to noise.
  This is particularly true for $\kin$, for which the non-axisymmetric
  part of $V_{\it los}$ is not much larger than typical measurement
  errors.  Merrifield \& Kuijken (1995) overcame this problem by
  projecting their slit spectra along the spatial direction thereby
  increasing the signal-to-noise ratio of $\kin$ significantly.  This
  projection amounts to carrying out the required velocity integral
  directly in photon space.  The signal in $\pin$ is typically much 
  better constrained, since bars tend to be bright.
  
\item{\it Sensitivity to errors in the disc's position angle.}  This
  was first recognized by Debattista \& Williams (2002).  They used
  two-dimensional Fabry-Perot absorption-line spectroscopy of NGC 7079
  to show that small errors in the derived position angle (PA) of the
  disc translate into large errors in $\om$.  For example, for NGC
  7079, they found that an error of as little as $5\degrees$ in the
  disc PA would result in an error of $100\%$ in $\om$.  Since the
  uncertainties in published values of PA for disc galaxies are often 
  at the $5\degrees$ level, each application of the TW method requires
  careful measurement of the disc PA.  Highly inclined galaxies
  and galaxies in which the bar is at about $45\degrees$ to the disc
  major axis, are less sensitive to errors in disc PA.  We note that
  past applications of the TW method using slits have been to well studied 
  galaxies (Merrifield \& Kuijken 1995; Gerssen \etal\ 1998), so that it
  is very unlikely that large errors of this type were introduced.  
\end{enumerate}

In this work, we have followed the prescriptions of Merrifield \& Kuijken
(1995) to deal with the centering and signal-to-noise problems.  We dealt
with the sensitivity to errors in disc PA by obtaining deep surface 
photometry of NGC 1023 in advance of the spectroscopy.

\section{Global properties of NGC 1023}
\label{sec:properties}

NGC 1023 (UGC 2154) is a highly-inclined lenticular galaxy classified
as SB0 by Nilson (1973), as SB0$_1$(5) by Sandage \& Tammann (1981),
and as SB0$^-$(rs) by de Vaucouleurs \etal\ (1991, hereafter RC3). Its
total $B$-band magnitude is $B_T=10.35$ (RC3), which, after correcting
for inclination and extinction, corresponds to $M_{B_T}^{0} = -19.96$
for an adopted distance of $10.2$ Mpc (Faber et al. 1997; $H_0=80$
\kmsm).

NGC 1023 is the brightest member of the LGG 70 group (Garcia 1993).  Its
closest companion, designated NGC 1023A by Hart et al. (1980), is a low 
luminosity condensation located near the eastern end of
the major axis. The centers of the two objects are separated by an angular
distance of $2\farcm7$ (RC3), corresponding to a projected linear
distance of about $8$ kpc at 10.2 Mpc. The proximity of
NGC 1023A to NGC 1023 led Arp (1966) to include NGC 1023 in the
section of his Atlas of Peculiar Galaxies (Plate 135) devoted to
E-like galaxies with nearby fragments. NGC 1023A was recognized as an
individual galaxy by Barbon \& Capaccioli (1975) and classified as
Magellanic irregular or late-type dwarf galaxy by Capaccioli, Lorenz
\& Afanasiev (1986).
High resolution radio observations by Sancisi \etal\ (1984) found a
complex structure and kinematics for the neutral hydrogen. Most of it
was shown to be outside of the galaxy and consists of tails and bridges
connecting NGC 1023 to three companions, including NGC 1023A. This
morphology suggests an interaction with NGC 1023A, a
conclusion also reached by Capaccioli \etal\ (1986). Although it is
rich in neutral hydrogen, NGC 1023 is undetected in ionized (Pogge \&
Eskridge 1993; S\'anchez-Portal et al. 2000) and molecular gas (Sofue
et al. 1993; Taniguchi et al. 1994), making application of the TW method
to it possible.

Surface photometry of NGC 1023 has been obtained in several optical
(\eg, Barbon \& Capaccioli 1975; Kormendy 1985; Lauer et al. 1995;
S\'anchez-Portal et al. 2000; Bower et al. 2001) and near-infrared
bands (M\"ollenhoff \& Heidt 2001). Detailed stellar kinematics have
been measured by Simien \& Prugniel (1997, $\rm
PA=87\degrees,177\degrees$), Neistein \etal\ (1999, $\rm
PA=87\degrees$), and recently by Bower \etal\ (2001, $\rm
PA=0\degrees,45\degrees,90\degrees$), who inferred the presence of a
supermassive black hole of $4 \times 10^7$ \msun\ by three-integral,
axisymmetric modelling. Sil'chenko (1999) observed the circumnuclear
region of NGC 1023 via two-dimensional spectroscopy, disentangling a
kinematically and chemically decoupled stellar disc (with a radius of
80 pc). The stars of this nuclear disc have a mean age of 7 Gyr and
are substantially younger than the rest of the galaxy, which is
characterized by extremely red integrated colours (RC3) and without
global star formation (Pogge \& Eskridge 1993).

\section{Surface Photometry and Data Reduction}
\label{sec:photometry}

\subsection{Observations and reduction}

We observed NGC 1023 during December 27-28, 2000 at the 1-m Jacobus
Kapteyn Telescope (JKT) located at the Roque de los Muchachos
Observatory (ORM), La Palma.  The detector consisted of a SITe2
device, with $2048\times2048$ pixels and an image scale of $0\farcs33$
pixel$^{-1}$, giving an unvignetted field of view of about
$10\arcmin\times10\arcmin$.  The seeing during the observing run
varied from $1\arcsec$ to $1\farcs5$.  Tab. \ref{tab:log_photometry}
gives the log of the photometric observations. The galaxy was imaged
using the Harris $B$, $V$ and $I$ bandpasses. Deep observations were
taken in all filters, reaching $\mu_I \simeq 22$ \mas\ and $\mu_B
\simeq 24$ \mas.

\begin{table}   
\caption{Log of the surface photometry observations}   
\begin{center}   
\begin{tabular}{l c c }   
\hline   
\multicolumn{1}{c}{Filter} &   
\multicolumn{1}{c}{Date} &     
\multicolumn{1}{c}{t$_{\it exp}$} \\  
\multicolumn{1}{c}{} &   
\multicolumn{1}{c}{} &      
\multicolumn{1}{c}{(s)} \\  
\hline   
$B$ & 27 Dec 2000 & $3\times900$ \\
    & 28 Dec 2000 & $900$ \\
$V$ & 28 Dec 2000 & $4\times480$ \\
$I$ & 27 Dec 2000 & $20\times180$ \\
    & 28 Dec 2000 & $10\times180$ \\
\hline   
\label{tab:log_photometry}   
\end{tabular}   
\end{center}   
\end{table}

The images were reduced using standard \iraf\footnote{\iraf~is
  distributed by NOAO, which is operated by AURA Inc., under contract
  with the National Science Foundation} tasks. A run of 10 bias images
was obtained each night, which were combined into a single bias frame
and subtracted from the object frames. The images were flat-fielded
using sky flats taken in all filters at the beginning and end of each
observing night. The sky background level was removed by fitting a
second order polynomial to the regions free of sources in the images.

For the photometric calibration of the galaxies, standard stars of
known magnitudes were observed. The calibration constant includes
corrections for atmospheric and Galactic extinction, and a colour
term. No attempt was made to correct for internal extinction. The
atmospheric extinction was taken from the differential aerosol
extinction for the ORM (King 1985). The Galactic extinction in the $B$-band 
was taken from Schlegel \etal\ (1998). We used the Galactic
extinction law from Cardelli \etal\ (1989) in order to get the extinction 
in the other filters.

\begin{figure}
\leavevmode{\psfig{figure=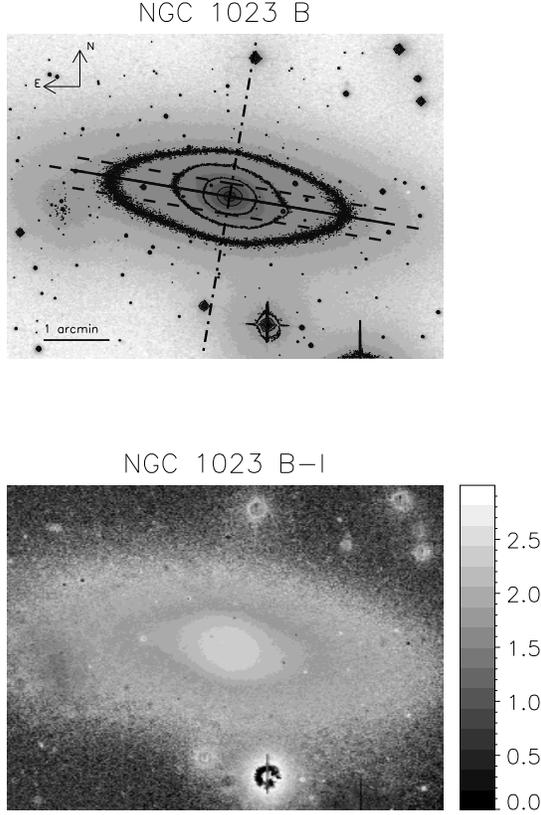,height=12.truecm,width=12.truecm}}
\caption[]{The $B$-band image (top) and $B-I$ colour map (bottom) of NGC 
  1023.  In the $B$-band image, the overplotted full line is the major
  axis (PA $=80\fdg2$), while the minor axis is indicated by the
  dot-dashed line.  The offset slits, parallel to the major axis and
  shifted by $12\arcsec$ to the north and $16\arcsec$ to the south,
  are shown by dashed lines.  The companion galaxy NGC 1023A can be
  seen to the east, just south of the major axis.  The contour levels
  shows are $\mu_B = 18$ to 22 \mas, at unit intervals, with the 
  outermost contour being $\mu_B = 22$ \mas.  The $B-I$ map shows
  that the strongest colour variations are associated with the bulge (which
  is red) and NGC 1023A (which is blue).}
\label{fig:maps}
\end{figure}

Fig. \ref{fig:maps} shows the reduced image of the galaxy in the
$B$-band and the $B-I$ colour map image.

\subsection{Isophotal analysis and bar length}

\begin{figure*}
\leavevmode{\psfig{figure=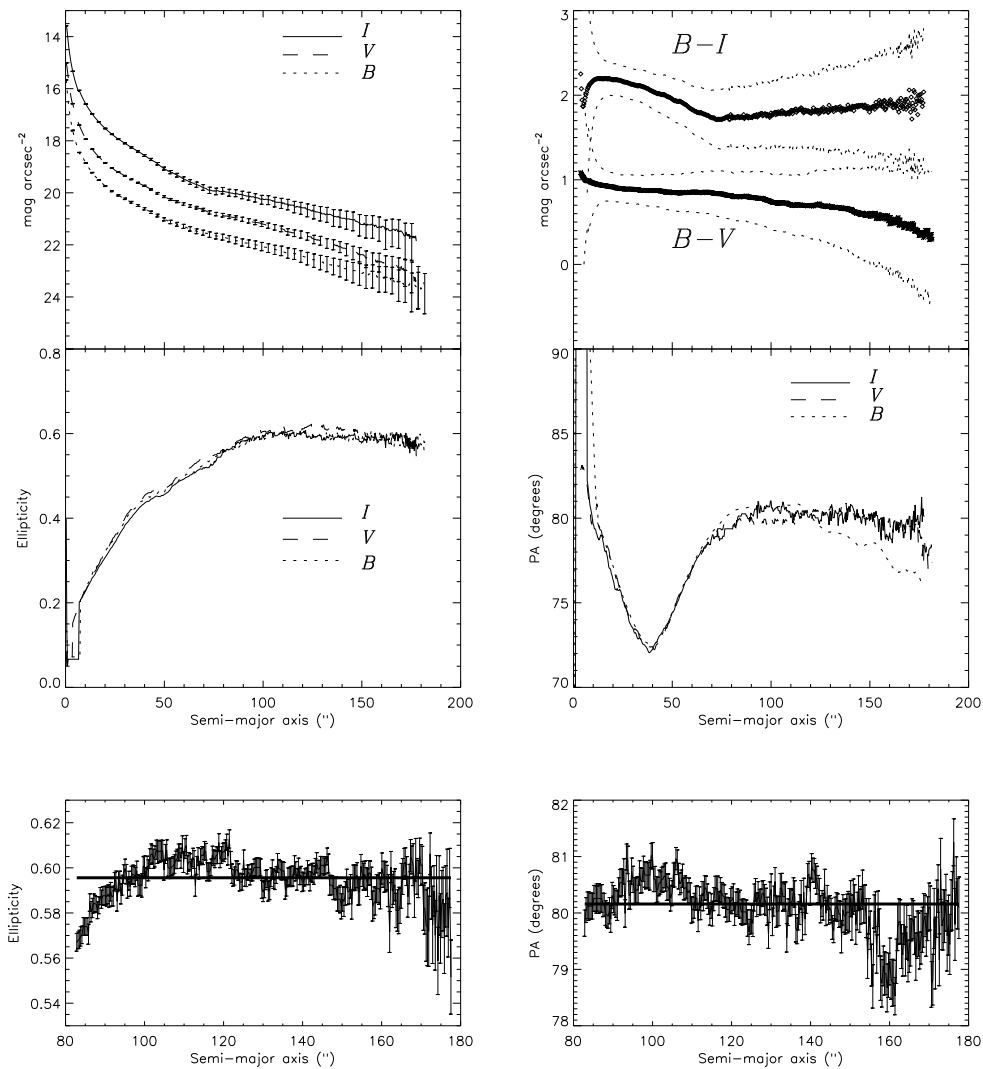,height=15.truecm,width=15.truecm}}
\caption[]{Surface brightness (top-left), colour (top-right), ellipticity
  (center-left) and PA (center-right) radial profiles of the ellipse
  fits to the isophotes of NGC 1023. The full, dotted and dashed lines
  show results for $I$, $V$ and $B$-bands respectively.  The bottom
  two panels show the $I-$band ellipticity (left) and PA (right) for
  $R \geq 80\arcsec$, which we averaged to determine the disc's
  inclination and PA.  The solid lines represent the average values in
  this range: $\epsilon = 0.595$ (corresponding to $i=66\fdg4$) and PA
  $=80\fdg2$.}
\label{fig:photometry}
\end{figure*}

We analysed the isophotal profiles of the galaxy by masking NGC 1023A and
the foreground stars, then fitting ellipses
using the \iraf\ task {\tt ELLIPSE}. We first allowed the centers of
the ellipses to vary, to test whether the optical disc is disturbed.
Within the errors of the fits, we found no evidence of a varying
center.  Thus the inner disc of NGC 1023 has had enough time to settle
if it had been disturbed in the tidal interaction suggested
by the \hi.  The ellipse fits were therefore repeated with the ellipse
centers fixed; the resulting azimuthally averaged surface brightness,
$B-I$ colour, ellipticity and PA profiles are plotted in Fig.
\ref{fig:photometry}.  These are in good agreement with the surface photometry
of S\'anchez-Portal \etal\ (2000), although our surface photometry reaches
fainter surface brightness.  The PA and ellipticity profiles are
similar for all bandpasses, suggesting that there is little, or
uniform, obscuration, as required for the TW method.  Furthermore, at
large radii, the $B-I$ profile (see Fig. \ref{fig:photometry}) is
flat, which is further evidence of little obscuration.  The only
structure seen in the $B-I$ profile is a small gradient in the region
where the bulge and bar dominate.  This gradient is in the sense of a
bulge redder than the disc, which is most likely caused by an age
difference between the bulge and disc stellar populations.  There is
also a small colour gradient in the bar region, but the color
variation is not patchy and is therefore unlikely to be caused by
dust.  We fitted the PA and inclination of the disc by averaging the
$I$-band data at semi-major axis greater than $80\arcsec$, finding PA 
$=80\fdg2 \pm 0\fdg5$ and $i=66\fdg4 \pm 1\fdg2$.  Application of the TW 
method generally requires a PA measured to $\sim 5\degrees$ accuracy 
(Debattista \& Williams 2002); we note that the value of the PA given in 
the RC3 is $87\degrees$.  Our value of the PA is in very good agreement 
with that measured by M\"ollenhoff \& Heidt (2001), who found PA $= 79\fdg15$ 
in the $J$-band, $80\fdg42$ in the $H$-band, and $78\fdg84$ in the
$K$-band, for an average of $79\fdg5 \pm 0\fdg7$.

We measured the disc exponential scale-length, $R_d$, from our
$I$-band surface photometry in the region outside the bar ($R \geq
80\arcsec$).  We obtained $R_d = 59\arcsec$, corresponding to 2.9 kpc
at our assumed distance.

\begin{figure*}
\leavevmode{\psfig{figure=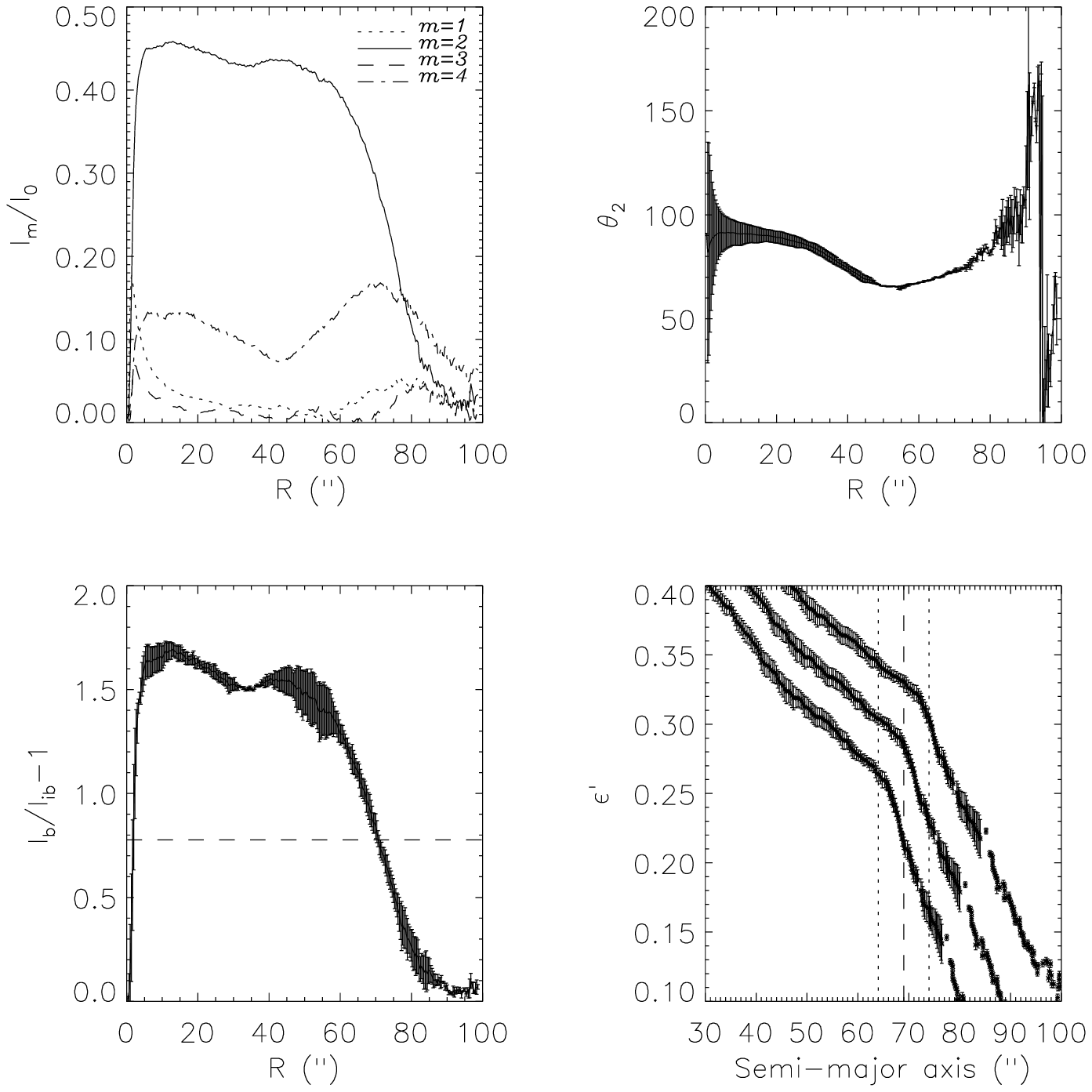,height=15.truecm,width=15.truecm}}
\caption[]{Fourier analysis of the deprojected $I$-band light distribution 
  of NGC 1023.  The top-left panel shows the relative amplitudes of
  the $m=1$ to $m=4$ Fourier components, which reveals that $m=2$
  dominates.  The top-right panel shows the phase angle of the $m=2$
  Fourier component; inside $R \simeq 30\arcsec$, the system is
  clearly dominated by the spheroidal bulge, which deprojects into a
  feature along the minor axis.  In the range $50\arcsec \ltsim R
  \ltsim 70\arcsec$, $\theta_2$ reaches a minimum of $\sim
  65\degrees$, but is not constant anywhere.  The panel on the
  bottom-left shows the bar/inter-bar intensity ratio.  The dashed
  line indicates one-half of the peak intensity (excluding the peak at
  $R < 30\arcsec$ since this is an artifact of the bulge
  deprojection).  The panel at bottom-right shows an alternative
  measure of $\len$, based on the deprojected ellipticities,
  $\epsilon^\prime$; the sharp break in $\epsilon^\prime$ at
  semi-major axis $\simeq 69\arcsec$ (indicated by the vertical dashed
  line) is identified as $\len$.  In this panel, the upper, central
  and lower set of points correspond to a deprojection assuming $i = $
  $65\fdg2$, $66\fdg4$ and $67\fdg6$ respectively.}
\label{fig:bar_fourier}
\end{figure*}

The most natural way to measure $\len$ is with a Fourier analysis of
the azimuthal luminosity profile (Ohta \etal\ 1990; Aguerri \etal\ 
2000), which we have applied here.  We began by deprojecting the
$I$-band image of the galaxy by a flux-conserving stretch along the
minor axis by the factor $1/\cos i$. NGC 1023 contains a significant
spheroidal bulge: the bulge-disc model decompositions of M\"ollenhoff
\& Heidt (2001) for this galaxy have a bulge that accounts for about
$30\%$ to $50\%$ of the near-infrared light.  Since we have not
attempted to subtract the bulge, we expect that, in deprojection,
it appears as a structure elongated with the minor axis, dominating
the center of the galaxy.  We then decomposed the deprojected
luminosity profile, $I(R,\phi)$, where $(R,\phi)$ are plane polar
coordinates in the galaxy frame, into a Fourier series:
\begin{equation}
I(R,\phi)=\frac{A_{0}(R)}{2}+ \sum_{m=1}^\infty (A_{m}(R) 
\cos(m\phi)+B_{m}(R)\sin(m\phi))
\end{equation}
where the coefficients are defined by:
\begin{equation}
A_{m}(R)=\frac{1}{\pi}\int^{2\pi}_{0}I(R,\phi)\cos(m\phi)d\phi
\end{equation}
and
\begin{equation}
B_{m}(R)=\frac{1}{\pi}\int^{2\pi}_{0}I(R,\phi)\sin(m\phi)d\phi
\end{equation}
The Fourier amplitude of the $m$-th component is defined as:
\begin{equation}
I_{m}(R) = \cases{ \displaystyle 
  A_{0}(R)/2 & $m = 0$ \cr
     \cr
 \sqrt{A_{m}^{2}(R)+B_{m}^{2}(R)} & $m \neq 0$ \cr}
\end{equation}
Fig. \ref{fig:bar_fourier} shows the first ($m=1,2,3,4$) relative
Fourier amplitudes, $I_{m}/I_{0}$, of the $I$-band image as a function
of $R$. The bar is evidenced by a strong $m=2$ component. Fig.
\ref{fig:bar_fourier} also shows the bar/inter-bar intensity as a
function of $R$. The bar intensity, $I_{b}$, is defined as the sum of
the even Fourier components, $I_{0}+I_{2}$, while the inter-bar
intensity, $I_{ib}$, is given by $I_{0}-I_{2}$ (Ohta \etal\ 1990;
Elmegreen \& Elmegreen 1990; Aguerri \etal\ 2000).  Ohta \etal\ (1990)
arbitrarily defined $\len$ as the outer radius for which $I_{b}/I_{ib}
= 2$; as Aguerri \etal\ (2000) pointed out, a fixed value of
$I_{b}/I_{ib}$ cannot account for the wide variety of bar luminosities
present in galaxies.  Instead, Aguerri \etal\ (2000) defined $\len$ as
the full width at half maximum (FWHM) of the curve of $I_{b}/I_{ib}$.
Using this definition, but excluding the peak in $I_{b}/I_{ib}$
resulting from the deprojected bulge, we obtain $\len = 69\arcsec \pm
5\arcsec$.  The phase angle, $\theta_{2}(R) =
\tan^{-1}(A_{2}(R)/B_{2}(R))$, of the $m=2$ component is shown in
Fig. \ref{fig:bar_fourier}.  The bulge dominates in the inner
$20\arcsec$ (with $\theta_{2}\approx 90 \degrees$, as expected), with
a transition from bulge to bar occurring in the range $20\arcsec
\ltsim R \ltsim 50\arcsec$. In the region $50\arcsec \ltsim R \ltsim
70\arcsec$, $\theta_{2}$ has a minimum, implying the bar extends to
somewhere here.  However, $\theta_{2}$ is not constant, because of the
influence of the bulge and the disc, and it cannot be used to
determine $\len$.

\begin{figure}
  \leavevmode{\psfig{figure=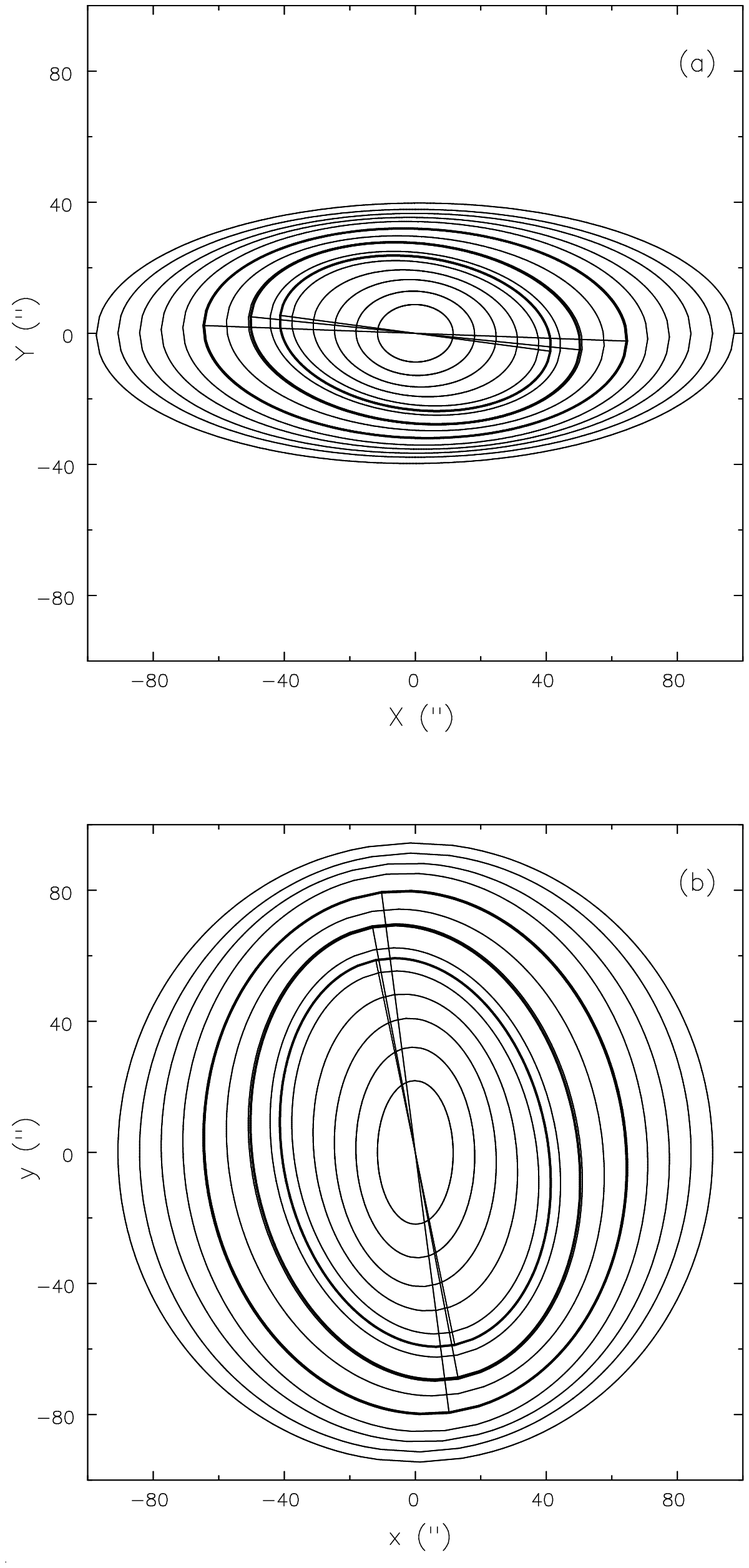,height=13.truecm,width=7.truecm}}
\caption[]{Best-fitting ellipses to the $I$-band isophotes (a) before and 
  (b) after deprojection to the NGC 1023 disc plane ($i = 66\fdg4$ and
  PA$=80\fdg2$).  The system has been rotated so that the disc major
  axis is horizontal.  Bold-line ellipses have deprojected semi-major
  axes of $60\arcsec$, $70\arcsec$ and $80\arcsec$; the straight lines
  indicate the major axes of these 3 ellipses.} 
\label{fig:ellipses}
\end{figure}

A independent measure of $\len$ was obtained by deprojecting the
ellipses which best fit the isophotes, averaging over the three bands,
and examining the deprojected ellipticity, $\epsilon^\prime$, as shown
in Fig.  \ref{fig:bar_fourier}.  We find a sharp break in the slope of
$\epsilon^\prime$ at $69\arcsec \pm 3\arcsec$, which we identify as
the end of the bar.  We repeated this experiment, deprojecting with
different $i$ and PA within the standard errors of these quantities.
The errors in PA do not significantly change the semi-major axis at
which the break occurs, while those in $i$ lead to changes in the
break semi-major axis of $\sim 3\arcsec$.  Thus, our second estimate
of $\len = 69\arcsec \pm 4\arcsec$.  Fig. \ref{fig:ellipses} plots the
projected and deprojected ellipses, from which the change at
semi-major axis $\sim 70\arcsec$ can be discerned.

Outside the inner $5\arcsec$, the largest twist due to the bar of the 
isophotes away from the sky-plane major axis is only $\simeq 8\degrees$, 
which occurs at $40\arcsec$.  We measured the bar's intrinsic angle from
the major axis, $\psi_{\rm bar}$, in the radial range $55\arcsec$ to 
$70\arcsec$, averaged over the 3 bands.  Due to the contamination from the 
bulge, the deprojected isophotal ellipses are not constant in this 
range, so we used as the uncertainty in $\psi_{\rm bar}$ the largest 
deviation from the average.  Thus we obtained 
$\psi_{\rm bar} = 102 \pm 2 \degrees$, so that the bar is only some 
$12\degrees$ from the minor axis, as can be seen in Fig. \ref{fig:ellipses}.  
This bar orientation, together with the large inclination and strong bulge, 
tend to obstruct our view of the bar.  In this sense, NGC 1023 is not an ideal 
candidate for the TW method, and the expected values of $\kin$ and $\pin$ are 
likely to be small, as we show in \S\ref{sec:pattern_speed}.  For this 
combination of inclination and $\psi_{\rm bar}$, Debattista (2002) 
suggests an accuracy of $\sim 2\fdg5$ in PA is required for accurate 
measurement of $\om$.

\section{Long-slit spectroscopy}
\label{sec:spectroscopy}

\subsection{Observations and data reduction}

The spectroscopic observations of NGC 1023 were carried out at the ORM
in La Palma, with the 3.6-m Telescopio Nazionale Galileo on February
01-02, 2001.
The telescope mounted the Low Resolution Spectrograph (DOLORES). The
HR-V grism No.~6 with 600 $\rm grooves\,mm^{-1}$ was used in
combination with the $0\farcs7\times8\farcm1$ slit and the thinned and
back-illuminated Loral CCD with $2048\,\times\,2048$ pixels of
$15\,\times\,15$ $\rm \mu m^2$. The wavelength range between 4697 and
6840 \AA\ was covered with a reciprocal dispersion of 1.055 \AA\ 
pixel$^{-1}$, which guarantees an adequate oversampling of the
instrumental broadening function. Indeed, the instrumental resolution,
obtained by measuring the width of emission lines of a comparison
spectrum after the wavelength calibration, was $3.10$ \AA\ (FWHM).
This corresponds to an instrumental dispersion $\sigma = 1.32$ \AA\ 
(\ie, $\sim80$ and $\sim60$ \kms\ at the blue and red edges of the
spectra, respectively). The angular sampling was $0\farcs275$
pixel$^{-1}$.

We obtained two 30-minutes spectra with the slit along the major axis
($\rm PA=80^\circ$) and two offset spectra with the slit parallel to
the major axis and shifted by $12\arcsec$ northward and $16\arcsec$
southward. The exposure time of each offset spectrum was 60 minutes.
During the observing run, we took spectra of the giant stars HR 2035
(G8III), HR 2429 (K1III), HR 2503 (K4III), HR 2701 (K0III), and HR
5370 (K3III) to use as templates in measuring the stellar kinematics.
At the beginning of each exposure, the slit was positioned by
acquiring with DOLORES a series of target images. A spectrum of the
comparison helium arc lamp was taken after each target exposure to
allow an accurate wavelength calibration.
The value of the seeing FWHM during the galaxy exposures ranged
between $0\farcs7$ and $1\farcs0$ as measured by fitting a
two-dimensional Gaussian to the guide star.

\begin{figure*}
 \leavevmode{\psfig{figure=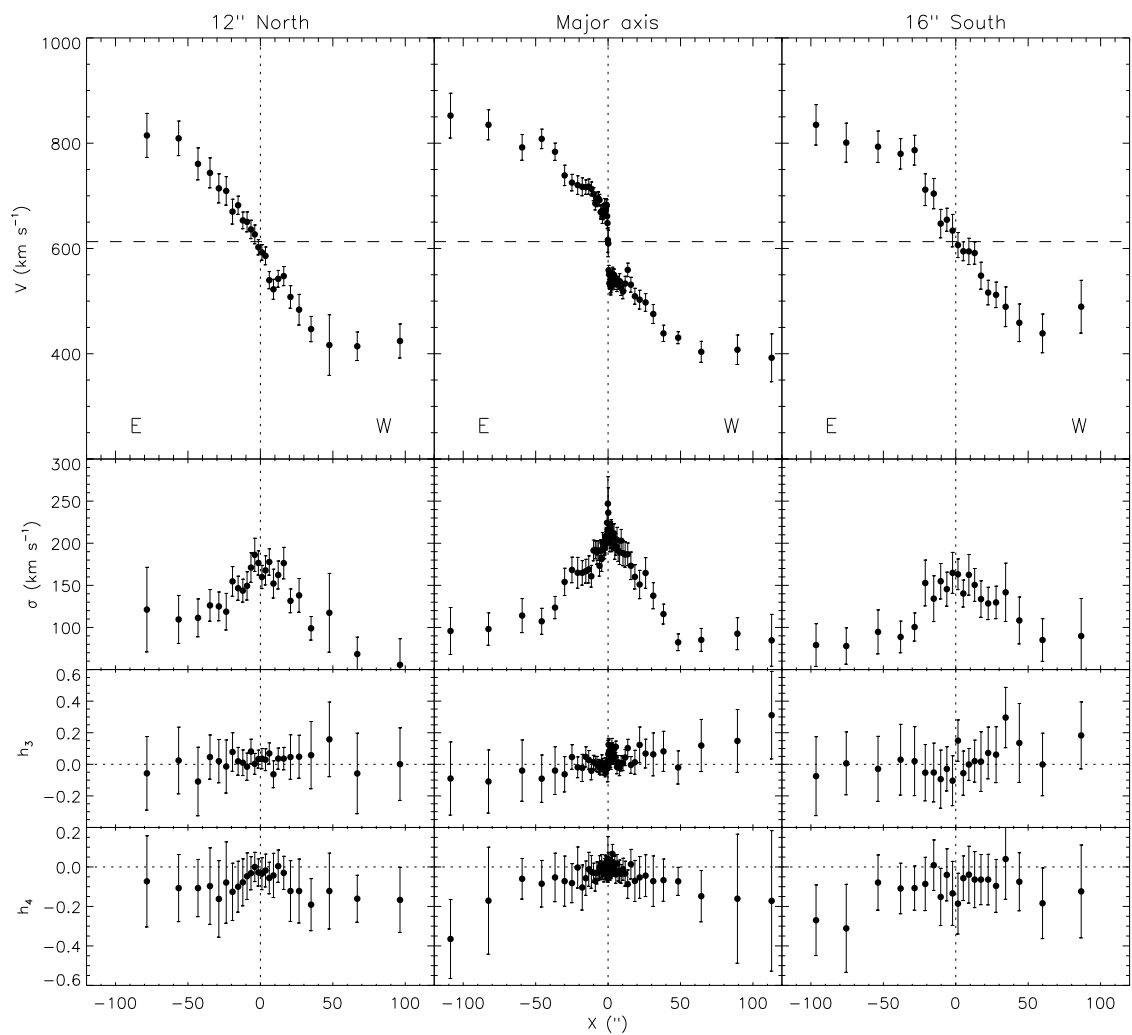,height=14.truecm,width=14.truecm}}
\vskip0.5truecm
\caption[]{Stellar kinematics of NGC 1023. Each column shows the radial 
  profiles of the four Gauss-Hermite moments ($V,\sigma,h_3$ and
  $h_4$) we measured along a given slit position.  In the top panels
  the horizontal dashed line indicates the heliocentric systemic
  velocity of the NGC 1023. All velocities are plotted as observed,
  without correcting for the inclination.  The full data are given in
  Tab. \ref{tab:kinematics}.}
\label{fig:kinematics}
\end{figure*}

All the spectra were bias subtracted, flatfield corrected, cleaned of
cosmic rays, and wavelength calibrated using standard
\midas\footnote{\midas\ is developed and maintained by the European
  Southern Observatory} routines.
The bias level was determined from the bias frames obtained during the
observing nights to check the CCD status. The flatfield correction was
performed by means of both quartz lamp and twilight sky spectra, which
were normalized and divided into all the spectra, to correct for
pixel-to-pixel sensitivity variations and large-scale illumination
patterns due to slit vignetting. Cosmic rays were identified by
comparing the counts in each pixel with the local mean and standard
deviation (as obtained from Poisson statistics by taking into account
the gain and read-out noise of the detector) and then corrected by
interpolating over. The residual cosmic rays were corrected by
manually editing the spectra.
The wavelength calibration was performed by means of the \midas\ 
package {\tt XLONG}. Each spectrum was rebinned using the wavelength
solution obtained from the corresponding arc-lamp spectrum.  We
checked that the wavelength rebinning had been done properly by
measuring the difference between the measured and predicted
wavelengths (Osterbrock \etal\ 1996) for the brightest night-sky
emission lines in the observed spectral range. The resulting accuracy 
in the wavelength calibration is $\sim1$ \kms.
The major-axis spectra were co-added using the center of the stellar
continuum as reference. In the resulting spectra, the contribution of
the sky was determined by interpolating along the outermost
$20\arcsec$ at the two edges of the slit, where the galaxy light was
negligible, and then subtracted.

\subsection{Stellar kinematics}

\begin{figure}
 \leavevmode{\psfig{figure=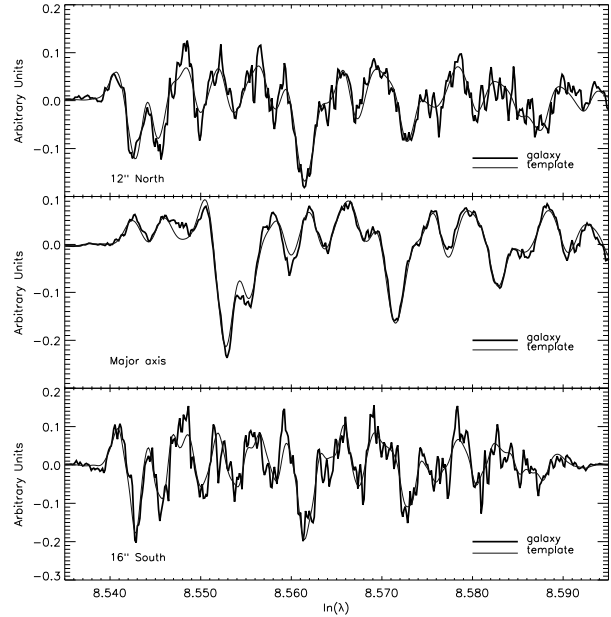,width=8.5truecm}}
\vskip0.5truecm
\caption[]{Galaxy spectra (thick lines), obtained by collapsing along the 
  spatial direction, of the major axis and the two offset positions.
  They are compared with the template spectra (thin line) convolved
  with their corresponding LOSVDs. Galaxy and template spectra have
  been continuum subtracted and tapered at the ends with a cosine bell
  function.} 
\label{fig:spectra}
\end{figure}

We measured the stellar kinematics from the galaxy absorption features
present in the wavelength range running from 4697 \AA\ to 5570 \AA\ 
and centered on the Mg line triplet ($\lambda\lambda\,5164,5173,5184$ 
\AA).
The spectra were rebinned along the spatial direction to obtain a
nearly constant signal-to-noise ratio larger than 20 per resolution
element (with a peak of 50 in the innermost regions of the major axis
spectrum). At the outermost radii the signal-to-noise ratio decreases
to $\sim10$. The galaxy continuum was removed row-by-row by fitting a
fourth to sixth order polynomial.

We used the Fourier Correlation Quotient method (FCQ, Bender 1990)
following the prescriptions of Bender, Saglia \& Gerhard (1994) and
adopting HR 2035 as kinematical template. This allowed us to derive,
for each spectrum, the line-of-sight velocity distribution (LOSVD)
along the slit and to measure its moments, namely the radial velocity
$v$, the velocity dispersion $\sigma$ and the values of the
coefficients $h_3$ and $h_4$.  At each radius, they have been derived
by fitting the LOSVD with a Gaussian plus third- and fourth-order
Gauss-Hermite polynomials $H_3$ and $H_4$, which describe the
asymmetric and symmetric deviations of the LOSVD from a pure Gaussian
profile (van der Marel \& Franx 1993; Gerhard 1993).
We derived errors on the LOSVD moments from photon statistics and CCD
read-out noise, calibrating them by Monte Carlo simulations as done
by Bender \etal\ (1994) and Bower \etal\ (2001), who have
recently derived the stellar kinematics of this galaxy along three
axes (different from ours).
In general, our errors are in the range of 10--20 \kms\ for $v$ and
$\sigma$, and of 0.03--0.1 for $h_3$ and $h_4$, becoming larger for
signal-to-noise ratios lower than $20$.  These errors do not take into
account possible systematic effects due to any template mismatch.  The
measured stellar kinematics along the major and the offset axes are
reported in Tab. \ref{tab:kinematics} and plotted in Fig.
\ref{fig:kinematics}.

\begin{table*}   
\caption{Stellar kinematics of NGC 1023. This data is plotted in Fig. 
\ref{fig:kinematics}.}   
\begin{tiny}
\begin{tabular}{rrrrr rrrrr}
\hline
\noalign{\smallskip}   
\multicolumn{1}{c}{$r$} &   
\multicolumn{1}{c}{$v$} &     
\multicolumn{1}{c}{$\sigma$} &   
\multicolumn{1}{c}{$h_{3}$} &   
\multicolumn{1}{c}{$h_{4}$} &
\multicolumn{1}{c}{$r$} &   
\multicolumn{1}{c}{$v$} &     
\multicolumn{1}{c}{$\sigma$} &   
\multicolumn{1}{c}{$h_{3}$} &   
\multicolumn{1}{c}{$h_{4}$} \\
\multicolumn{1}{c}{[$''$]} &   
\multicolumn{1}{c}{[km s$^{-1}$]} &      
\multicolumn{1}{c}{[km s$^{-1}$]} &   
\multicolumn{1}{c}{} &  
\multicolumn{1}{c}{} &
\multicolumn{1}{c}{[$''$]} &   
\multicolumn{1}{c}{[km s$^{-1}$]} &      
\multicolumn{1}{c}{[km s$^{-1}$]} &   
\multicolumn{1}{c}{} &  
\multicolumn{1}{c}{} \\
\noalign{\smallskip}   
\hline  
\noalign{\smallskip}   
         &               &               &                 &                  & $  25.89$& $497.4\pm16.8$& $164.7\pm18.2$& $ 0.068\pm0.091$& $-0.044\pm0.106$\\
\multicolumn{5}{c}{Major Axis}                                                & $  31.20$& $475.4\pm18.1$& $137.7\pm15.5$& $ 0.063\pm0.136$& $-0.072\pm0.128$\\
         &               &               &                 &                  & $  38.25$& $438.6\pm15.5$& $115.9\pm11.8$& $ 0.082\pm0.127$& $-0.067\pm0.107$\\
$-108.76$& $852.3\pm42.5$& $ 95.8\pm27.9$& $-0.090\pm0.232$& $-0.365\pm0.200$ & $  48.38$& $430.4\pm11.4$& $ 82.4\pm 9.9$& $-0.020\pm0.105$& $-0.073\pm0.071$\\
$ -82.56$& $834.8\pm28.5$& $ 98.1\pm19.2$& $-0.109\pm0.200$& $-0.171\pm0.271$ & $  64.33$& $403.5\pm19.9$& $ 85.2\pm13.6$& $ 0.119\pm0.165$& $-0.148\pm0.130$\\
$ -59.33$& $791.9\pm24.5$& $114.1\pm20.2$& $-0.040\pm0.194$& $-0.060\pm0.103$ & $  89.29$& $407.5\pm28.0$& $ 92.6\pm19.0$& $ 0.148\pm0.199$& $-0.161\pm0.327$\\
$ -45.82$& $808.1\pm18.5$& $107.3\pm15.4$& $-0.091\pm0.150$& $-0.085\pm0.118$ & $ 112.90$& $392.2\pm45.5$& $ 84.7\pm30.8$& $ 0.311\pm0.277$& $-0.172\pm0.356$\\
$ -36.62$& $783.6\pm16.4$& $123.6\pm13.2$& $-0.040\pm0.151$& $-0.053\pm0.123$ &          &               &               &                 &                 \\
$ -29.99$& $738.7\pm19.4$& $154.1\pm16.2$& $-0.063\pm0.112$& $-0.072\pm0.127$ & \multicolumn{5}{c}{$12''$ North Offset}\\
$ -24.93$& $725.0\pm15.7$& $168.3\pm15.2$& $ 0.046\pm0.078$& $-0.082\pm0.099$ &          &               &               &                 &                 \\
$ -20.98$& $720.3\pm17.7$& $164.9\pm18.3$& $-0.019\pm0.066$& $-0.003\pm0.104$ & $ -78.35$& $814.6\pm41.8$& $121.2\pm50.2$& $-0.057\pm0.233$& $-0.073\pm0.231$\\
$ -17.83$& $717.0\pm17.0$& $164.8\pm14.6$& $-0.023\pm0.086$& $-0.104\pm0.114$ & $ -56.44$& $809.1\pm32.8$& $109.5\pm28.5$& $ 0.024\pm0.211$& $-0.107\pm0.170$\\
$ -15.37$& $716.4\pm13.7$& $166.8\pm14.1$& $ 0.047\pm0.064$& $-0.057\pm0.087$ & $ -43.07$& $760.6\pm30.4$& $111.4\pm22.5$& $-0.109\pm0.217$& $-0.107\pm0.145$\\
$ -13.32$& $717.4\pm14.4$& $168.9\pm16.1$& $ 0.028\pm0.059$& $-0.012\pm0.086$ & $ -34.80$& $743.5\pm28.5$& $126.2\pm18.8$& $ 0.046\pm0.140$& $-0.097\pm0.194$\\
$ -11.53$& $713.1\pm13.1$& $160.8\pm12.9$& $-0.041\pm0.057$& $-0.025\pm0.083$ & $ -28.68$& $714.1\pm27.6$& $125.0\pm17.1$& $ 0.020\pm0.137$& $-0.162\pm0.194$\\
$ -10.02$& $703.1\pm12.4$& $191.4\pm12.3$& $ 0.014\pm0.043$& $-0.029\pm0.048$ & $ -23.61$& $709.2\pm27.1$& $118.7\pm21.7$& $-0.014\pm0.168$& $-0.079\pm0.206$\\
$  -8.79$& $685.3\pm12.2$& $192.1\pm10.5$& $ 0.007\pm0.043$& $-0.071\pm0.051$ & $ -19.23$& $669.9\pm23.6$& $154.7\pm17.6$& $ 0.078\pm0.122$& $-0.126\pm0.145$\\
$  -7.69$& $693.1\pm13.9$& $189.1\pm14.0$& $ 0.002\pm0.048$& $-0.030\pm0.056$ & $ -15.39$& $682.0\pm17.3$& $146.6\pm14.3$& $ 0.019\pm0.088$& $-0.100\pm0.129$\\
$  -6.74$& $695.1\pm12.2$& $191.3\pm11.7$& $-0.010\pm0.041$& $-0.024\pm0.045$ & $ -12.11$& $653.2\pm15.9$& $143.7\pm13.6$& $ 0.010\pm0.081$& $-0.078\pm0.119$\\
$  -5.91$& $691.8\pm11.0$& $173.2\pm12.2$& $-0.019\pm0.043$& $-0.001\pm0.058$ & $  -9.22$& $650.4\pm19.0$& $149.4\pm16.8$& $-0.015\pm0.086$& $-0.047\pm0.118$\\
$  -5.08$& $668.9\pm12.9$& $191.4\pm12.8$& $-0.024\pm0.043$& $-0.017\pm0.048$ & $  -6.49$& $636.1\pm17.5$& $171.1\pm17.5$& $ 0.081\pm0.078$& $-0.032\pm0.084$\\
$  -4.40$& $670.7\pm13.0$& $181.4\pm14.9$& $ 0.013\pm0.047$& $ 0.004\pm0.059$ & $  -3.88$& $626.6\pm17.8$& $186.2\pm19.7$& $ 0.003\pm0.066$& $ 0.000\pm0.074$\\
$  -3.85$& $660.2\pm12.6$& $200.7\pm12.1$& $-0.005\pm0.042$& $-0.022\pm0.042$ & $  -1.40$& $602.2\pm14.5$& $176.6\pm14.3$& $ 0.033\pm0.057$& $-0.028\pm0.069$\\
$  -3.30$& $665.0\pm10.6$& $200.2\pm 9.9$& $-0.032\pm0.035$& $-0.013\pm0.035$ & $   1.07$& $592.4\pm14.7$& $159.9\pm13.8$& $ 0.036\pm0.064$& $-0.034\pm0.081$\\
$  -2.75$& $669.6\pm11.5$& $196.0\pm10.3$& $-0.022\pm0.039$& $-0.036\pm0.040$ & $   3.53$& $585.8\pm16.9$& $167.8\pm17.3$& $ 0.029\pm0.069$& $-0.018\pm0.083$\\
$  -2.34$& $681.8\pm12.7$& $194.3\pm12.9$& $-0.028\pm0.041$& $-0.003\pm0.045$ & $   6.13$& $539.7\pm16.3$& $177.7\pm15.6$& $ 0.069\pm0.067$& $-0.056\pm0.080$\\  
$  -2.07$& $674.1\pm11.3$& $198.6\pm12.5$& $ 0.004\pm0.037$& $ 0.019\pm0.041$ & $   9.02$& $522.3\pm19.0$& $152.1\pm17.1$& $-0.063\pm0.085$& $-0.043\pm0.112$\\  
$  -1.79$& $672.3\pm11.8$& $199.6\pm11.1$& $-0.040\pm0.039$& $-0.019\pm0.039$ & $  12.29$& $542.2\pm15.9$& $162.3\pm16.6$& $ 0.036\pm0.067$& $ 0.004\pm0.082$\\  
$  -1.52$& $674.0\pm10.3$& $203.1\pm10.7$& $-0.030\pm0.033$& $ 0.017\pm0.035$ & $  16.12$& $547.4\pm18.2$& $176.4\pm18.7$& $ 0.036\pm0.071$& $-0.030\pm0.084$\\  
$  -1.24$& $683.3\pm11.3$& $208.2\pm10.2$& $-0.027\pm0.038$& $-0.029\pm0.035$ & $  20.76$& $507.8\pm21.4$& $131.6\pm14.1$& $ 0.046\pm0.141$& $-0.122\pm0.154$\\  
$  -0.97$& $681.2\pm11.2$& $202.6\pm10.5$& $-0.030\pm0.037$& $-0.016\pm0.036$ & $  26.72$& $483.6\pm29.1$& $138.1\pm20.0$& $ 0.048\pm0.136$& $-0.122\pm0.162$\\  
$  -0.69$& $661.3\pm12.2$& $224.3\pm11.1$& $-0.008\pm0.042$& $-0.029\pm0.037$ & $  34.99$& $446.8\pm24.0$& $ 99.0\pm14.0$& $ 0.058\pm0.213$& $-0.191\pm0.132$\\  
$  -0.42$& $648.1\pm10.5$& $216.1\pm10.5$& $-0.024\pm0.034$& $ 0.005\pm0.033$ & $  47.61$& $416.4\pm57.6$& $117.4\pm46.8$& $ 0.158\pm0.236$& $-0.122\pm0.192$\\  
$  -0.14$& $615.9\pm23.2$& $247.0\pm32.0$& $ 0.020\pm0.131$& $ 0.041\pm0.113$ & $  66.86$& $414.1\pm27.1$& $ 68.4\pm20.1$& $-0.058\pm0.255$& $-0.161\pm0.119$\\  
$   0.13$& $609.5\pm25.3$& $236.3\pm29.6$& $ 0.064\pm0.100$& $-0.002\pm0.076$ & $  96.37$& $424.1\pm32.3$& $ 55.7\pm31.0$& $ 0.001\pm0.230$& $-0.167\pm0.165$\\  
$   0.41$& $558.6\pm 9.2$& $215.1\pm10.2$& $ 0.124\pm0.031$& $-0.002\pm0.036$ &         &               &               &                 &                 \\  
$   0.68$& $550.7\pm 9.4$& $215.5\pm 9.7$& $ 0.082\pm0.032$& $-0.001\pm0.033$ & \multicolumn{5}{c}{$16''$ South Offset}\\
$   0.96$& $534.0\pm10.6$& $211.9\pm10.7$& $ 0.062\pm0.037$& $-0.011\pm0.037$ &         &               &               &                 &                 \\  
$   1.23$& $532.8\pm10.1$& $200.7\pm10.8$& $ 0.066\pm0.036$& $-0.011\pm0.040$ & $ -96.43$& $834.8\pm38.4$& $ 79.1\pm25.4$& $-0.075\pm0.250$& $-0.270\pm0.179$\\  
$   1.51$& $536.2\pm10.1$& $205.6\pm11.3$& $ 0.083\pm0.035$& $ 0.003\pm0.039$ & $ -75.58$& $800.9\pm37.1$& $ 77.9\pm21.6$& $ 0.006\pm0.199$& $-0.311\pm0.223$\\  
$   1.78$& $527.2\pm11.7$& $211.4\pm10.9$& $ 0.019\pm0.040$& $-0.028\pm0.038$ & $ -53.66$& $793.2\pm29.8$& $ 94.7\pm26.1$& $-0.029\pm0.206$& $-0.079\pm0.140$\\  
$   2.06$& $527.3\pm15.7$& $211.5\pm16.4$& $ 0.020\pm0.053$& $-0.006\pm0.052$ & $ -38.07$& $779.7\pm28.9$& $ 88.7\pm18.8$& $ 0.029\pm0.224$& $-0.109\pm0.128$\\  
$   2.33$& $529.1\pm12.2$& $213.2\pm12.0$& $ 0.052\pm0.042$& $-0.012\pm0.040$ & $ -28.42$& $786.6\pm28.3$& $100.5\pm16.7$& $ 0.020\pm0.218$& $-0.107\pm0.112$\\  
$   2.61$& $534.5\pm13.6$& $206.3\pm16.4$& $ 0.116\pm0.048$& $ 0.019\pm0.056$ & $ -21.04$& $711.6\pm30.2$& $152.8\pm27.2$& $-0.053\pm0.178$& $-0.086\pm0.135$\\  
$   3.01$& $552.2\pm12.0$& $207.6\pm15.6$& $ 0.097\pm0.039$& $ 0.066\pm0.048$ & $ -15.18$& $704.1\pm28.8$& $134.2\pm27.2$& $-0.052\pm0.186$& $ 0.009\pm0.128$\\  
$   3.56$& $550.0\pm11.7$& $210.8\pm12.0$& $ 0.040\pm0.040$& $-0.004\pm0.040$ & $ -10.40$& $647.1\pm27.0$& $154.9\pm21.1$& $-0.094\pm0.184$& $-0.152\pm0.145$\\  
$   4.11$& $549.5\pm11.3$& $211.7\pm11.4$& $ 0.058\pm0.039$& $-0.009\pm0.039$ & $  -6.15$& $654.5\pm21.9$& $145.5\pm20.2$& $-0.029\pm0.140$& $-0.040\pm0.132$\\  
$   4.66$& $533.6\pm11.7$& $194.4\pm13.4$& $ 0.046\pm0.041$& $ 0.009\pm0.047$ & $  -2.18$& $633.8\pm30.8$& $164.8\pm24.3$& $-0.104\pm0.158$& $-0.134\pm0.162$\\  
$   5.35$& $538.3\pm 9.0$& $196.4\pm10.7$& $ 0.111\pm0.036$& $-0.038\pm0.045$ & $   1.52$& $606.4\pm23.7$& $163.1\pm18.2$& $ 0.150\pm0.131$& $-0.186\pm0.154$\\  
$   6.17$& $531.4\pm13.4$& $204.0\pm14.8$& $-0.017\pm0.043$& $ 0.017\pm0.046$ & $   5.23$& $594.6\pm18.0$& $140.3\pm16.0$& $-0.056\pm0.140$& $-0.057\pm0.113$\\  
$   7.00$& $534.5\pm12.6$& $191.2\pm12.8$& $ 0.013\pm0.043$& $-0.022\pm0.049$ & $   9.06$& $594.3\pm24.9$& $162.5\pm24.2$& $-0.001\pm0.101$& $-0.039\pm0.144$\\  
$   7.95$& $537.8\pm14.2$& $189.2\pm14.5$& $-0.024\pm0.048$& $-0.021\pm0.056$ & $  13.05$& $591.2\pm21.2$& $150.5\pm19.5$& $ 0.020\pm0.133$& $-0.064\pm0.142$\\  
$   9.06$& $525.0\pm14.3$& $203.0\pm13.4$& $ 0.006\pm0.048$& $-0.035\pm0.047$ & $  17.43$& $548.2\pm25.6$& $133.7\pm21.7$& $ 0.017\pm0.189$& $-0.064\pm0.127$\\  
$  10.29$& $518.0\pm13.7$& $188.2\pm14.9$& $ 0.007\pm0.047$& $-0.012\pm0.056$ & $  22.37$& $516.1\pm23.3$& $128.5\pm19.2$& $ 0.072\pm0.164$& $-0.064\pm0.128$\\  
$  11.80$& $533.2\pm13.9$& $186.6\pm15.0$& $ 0.043\pm0.051$& $-0.034\pm0.063$ & $  27.83$& $511.8\pm24.2$& $129.7\pm19.3$& $ 0.061\pm0.177$& $-0.096\pm0.134$\\  
$  13.58$& $559.2\pm12.8$& $186.9\pm13.4$& $ 0.103\pm0.055$& $-0.087\pm0.072$ & $  34.42$& $489.1\pm37.7$& $141.7\pm34.6$& $ 0.296\pm0.191$& $ 0.040\pm0.204$\\  
$  15.77$& $531.0\pm14.1$& $173.2\pm16.3$& $-0.004\pm0.053$& $ 0.014\pm0.075$ & $  43.89$& $458.7\pm35.7$& $108.3\pm27.8$& $ 0.135\pm0.250$& $-0.075\pm0.147$\\  
$  18.50$& $509.1\pm15.1$& $160.1\pm14.4$& $ 0.013\pm0.076$& $-0.071\pm0.104$ & $  59.92$& $438.6\pm36.8$& $ 85.1\pm25.3$& $-0.001\pm0.198$& $-0.184\pm0.179$\\  
$  21.79$& $502.6\pm17.8$& $151.0\pm16.8$& $ 0.123\pm0.113$& $-0.053\pm0.106$ & $  86.58$& $489.2\pm50.1$& $ 89.8\pm44.6$& $ 0.183\pm0.212$& $-0.124\pm0.235$\\  
\noalign{\smallskip}   
\hline 
\label{tab:kinematics}
\end{tabular}   
\end{tiny}
\end{table*}

\section{Pattern speed}
\label{sec:pattern_speed}

\begin{figure*}
 \leavevmode{\psfig{figure=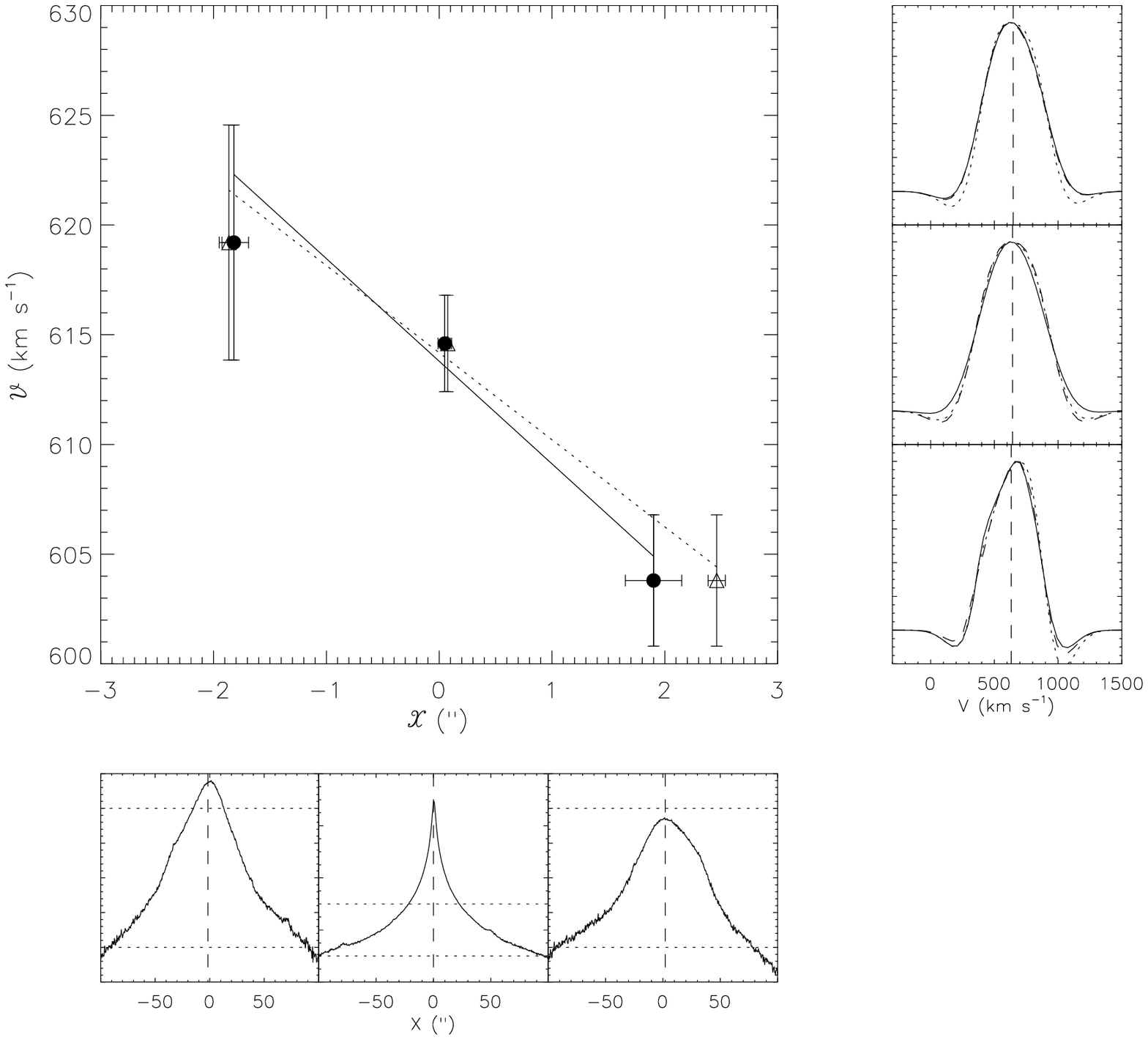,height=15.truecm,width=15.truecm}}
\vskip0.5truecm
\caption[]{The pattern speed measurement for NGC 1023.  The bottom 3 panels
  show the $V$-band profiles for the three slits ($12\arcsec$N, major
  axis and $16\arcsec$S from left to right) with $\pin$ indicated by
  the dashed vertical lines.  The horizontal dotted lines indicate
  $\mu_V = 19$ \mas\ and $\mu_V = 21$ \mas.  The 3 panels on the right
  indicate the LOSVD for the three slits ($12\arcsec$N, major axis and
  $16\arcsec$S from top to bottom), with $\kin$ indicated by the
  dashed lines.  The LOSVDs are constructed from the Gauss-Hermite
  moments; the 3 LOSVDs in each of these panels were obtained from
  different wavelength ranges (here, each is renormalized to the same
  maximum value, for clarity).  The central panel plots $\kin$ against
  $\pin$ (solid circles); the slope of the best fit straight line
  (indicated by the solid line) is $\om \sin i = 4.7 \pm 1.7$ \kmsa.
  The open triangles and dotted line indicates the results for using
  just the $I$-band to measure $\pin$, which gives the largest values of 
  $|\pin|$ and therefore the smallest slope, $\om \sin i = 4.0 \pm 1.4$ \kmsa.}
\label{fig:pattern_speed}
\end{figure*}

To compute the mean position of stars, $\pin$, along the slits, we
extracted profiles from the $B$, $V$ and $I$-band surface photometry along the
positions of the slits.  The profiles in $V$-band (within which the
wavelength range used in the spectroscopy falls) match very well the
profiles obtained by collapsing the spectra along the wavelength
direction, confirming that the slits were placed as intended.  We used 
the broad band profiles to compute $\pin$ because
these are less noisy than the spectral profiles, particularly at large
radii.  We therefore computed three values of $\pin$ at each slit
position, $\pin_B$, $\pin_V$ and $\pin_I$.  Each value of $\pin$ was
computed by Monte Carlo simulation, with photon, read-out and sky
noise to compute the errors.  Another advantage to using the broad
band surface photometry is that it allows us to test for contamination by 
NGC 1023A to the integrals.  Formally, the integrals
in Eqn.  \ref{eq:TW_equation} are over $-\infty \leq X \leq \infty$,
but can be limited to $-X_{max} \leq X \leq X_{max}$ if $X_{max}$ has
reached the axisymmetric part of the disc; still larger values of 
$X_{max}$ add noise only.  In the case of NGC 1023,
the presence of NGC 1023A introduces some additional light at negative
$X$.  The most severely contaminated slit is the one at $16\arcsec$S,
where $\pin$ continues changing as ever larger values of $X_{max}$ are
used.  We therefore found it useful to compare the $16\arcsec$S
profiles to the profiles at $16\arcsec$N.  For these latter profiles,
there is no further signal to $\pin$ beyond $X_{max} = 60\arcsec$.  The
comparison also showed that there is little or no contamination to the
$16\arcsec$S slit inside $X_{max} \sim 55\arcsec$.  Furthermore, since NGC
1023A is quite blue, the contamination decreases in going from $B$ to
$V$, and especially to $I$ band.  Therefore, to measure $\pin$ at
$16\arcsec$S, we averaged the Monte Carlo results between $62\arcsec
\leq X_{max} \leq 75\arcsec$ ($75\arcsec \leq R \leq 85\arcsec$, where the
disc is roughly axisymmetric) and then averaged over each band.  In
the remaining slits, contamination by NGC 1023A in the major axis
begins at $R \geq 100\arcsec$ in all bands and is easily avoided,
while the $12\arcsec$N slit has no obvious contamination.  We
therefore also averaged them over $75\arcsec \leq R \leq 85\arcsec$.
Because of the color gradient in the inner $\sim 70\arcsec$ (Fig.
\ref{fig:photometry}), the value of $\pin$ in the offset slits changes
with the color band used, with $|\pin_I| > |\pin_V| \gtsim |\pin_B|$.
Therefore, by averaging over bands and then taking the largest
discrepancy from the average as our error value, we are confident that
we take into account any difference between the narrow wavelength
range of the spectroscopy and the broad one of the surface photometry.  The
values thus obtained are given in Tab.  \ref{tab:TW_values}.

If we have successfully avoided contamination by NGC 1023A, then we expect
that, since NGC 1023 is nearly point-symmetric about its center, the profile
extracted from the surface photometry at $Y$ should have a value of $\pin$
roughly the negative of value for the profile extracted at $-Y$.  For a
profile extracted from the surface photometry at $16\arcsec$N, we found 
$\pin = -2\farcs31 \pm 0\farcs15$ averaged over the 3 bands, which is 
within $2\sigma$ of our value for $-\pin$ at $16\arcsec$S.

To measure the luminosity-weighted line-of-sight stellar velocity,
$\kin$, for each slit position, we collapsed each bi-dimensional
spectrum along its spatial direction to obtain a one-dimensional
spectrum. The resulting spectra have been analysed with the FCQ method
using HR 2035 as template star. $\kin$ is the radial velocity
derived from the LOSVD of the one-dimensional spectra (see Fig.
\ref{fig:spectra}).  For each slit position, the uncertainties on $\kin$ were 
estimated by measuring it in different wavelength ranges between 4910 \AA\ 
and 5560 \AA\ (but always including the Mg triplet).  The $\kin$ values we 
derived along each slit are given in Tab. \ref{tab:TW_values}.
We integrated the spectral rows at $X_{max} = 140\arcsec$,
$100\arcsec$ and $80\arcsec$ in the major-axis, northern and southern
offset spectra, respectively. For the major-axis and northern offset
spectra we were limited by the noise, which becomes dominant at larger
radii ($S/N\leq3$), while for the southern offset spectrum we decided
to also try $X_{max} = 60\arcsec$, to check for contamination by NGC 1023A.
We found that different choices of $X_{max}$ do not affect our results for
$\kin$.  The values of $\kin$ we derived by integrating both offset spectra 
out to $X_{max} = 140\arcsec$ (after removing by linear interpolation the 
contribution of foreground stars) are close to those in Tab. 
\ref{tab:TW_values}, although their errors increase by a factor 1.6 and 
2 along the north and south slit, respectively.  Therefore, within
the errors obtained from the varying wavelength range, there is no
evidence of contamination by NGC 1023A in $\kin$ for the southern
offset slit.  

\begin{table}
\centering
\caption{The values of $\kin$ and $\pin$ for the 3 slits.}
\begin{tabular}{lcc} 
\hline 
Offset & $\pin$      & $\kin$ \\
       & ($\arcsec$) & (\kms) \\
\hline 
$12\arcsec$ N & $-1.82 \pm 0.13$ & $619.1 \pm 5.4$  \\
Major axis    & $+0.05 \pm 0.06$ & $614.5 \pm 2.2$  \\
$16\arcsec$ S & $+1.90 \pm 0.25$ & $603.7 \pm 3.0$  \\
\hline
\label{tab:TW_values}
\end{tabular}
\end{table}

By performing our analysis with all the remaining template spectra, we
found that $\kin$ values are not affected by template mismatch (at
least for our set of template stars). This is in agreement with the
results of Bender \etal\ (1994) on the determination of the radial
velocities with FCQ.

Using the data of Tab. \ref{tab:TW_values}, in Fig.
\ref{fig:pattern_speed} we plot $\kin$ versus $\pin$, fitting a straight 
line (using subroutine {\tt fitexy} in Numerical Recipes).  The slope of 
this line is $\om \sin i = 4.7 \pm 1.7$ \kmsa, which gives $\om = 5.1 \pm 1.8$ 
\kmsa\ (corresponding to $\om = 104 \pm 34$ \kmsk\ at 10.2 Mpc).  We have 
also fitted a ``slowest-bar'' line, using the values of $\pin_I$ (which are 
largest, and therefore give a shallower slope) in the $12\arcsec$N and 
major-axis, and adopting the value of $\pin_I$ from $16\arcsec$N, which gives
a slightly larger $|\pin|$ than the $16\arcsec$S profile.  The slope 
of this line, now ignoring the uncertainties in $\pin$, is 
$\om \sin i = 4.0 \pm 1.4$ \kmsa\ ($\om = 4.3 \pm 1.5$ \kmsa).

\begin{figure}
 \leavevmode{\psfig{figure=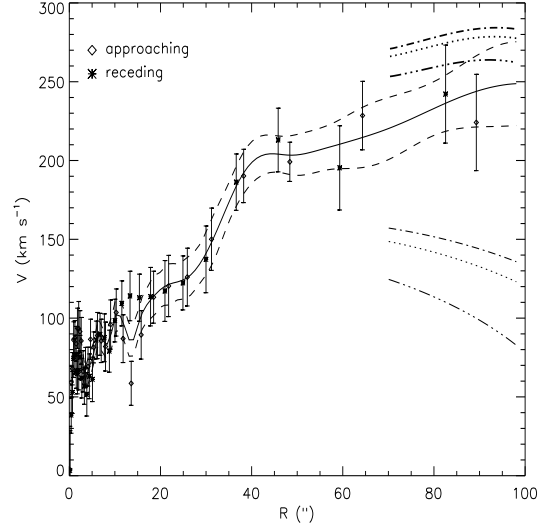,height=7.5truecm,width=7.5truecm}}
\caption[]{The stellar velocity streaming curve, $V_*$, derived by 
 subtracting $V_{\it sys}$, folding and deprojecting the observed velocities 
  on the major axis.  The data from the major axis are shown in diamonds
  (approaching side) and stars (receding side).  The solid line and
  the dashed lines flanking it are our fitted spline together with the
  $1\sigma$ error interval.  The asymmetric drift terms ($\sqrt{V_c^2
    - V_*^2}$) are shown by the dot-dashed, dotted, and dash
  triple-dotted lines ($\alpha = 0.0$, $0.5$, and $1.0$ respectively)
  in the lower-right corner at $R \geq 70\arcsec$.  The corresponding
  circular rotation curves, $V_c$, are shown in the bold lines.} 
\label{fig:circular_velocity}
\end{figure}

To then determine $\vpd$, we need to measure $\lag$, which we
approximate by the co-rotation radius, $R_c$, from the axisymmetric
approximation.  Debattista (1998) found that, for strong bars in
$N$-body simulations, the approximation $\lag \simeq R_c$ involves an
error of $\sim 5\%$.  We therefore measured the rotation curve,
$V_c(R)$, of NGC 1023.  This is a two step process: first we fitted
tilted rings (with constant PA and inclination fixed to the value
obtained from the surface photometry) to the data from all our slits,
interpolating with cubic splines where necessary.  This allowed us to
measure the heliocentric systemic velocity, $V_{\it sys} = 613\pm3$
\kms, which is compared to the values based on optical data available
in the literature in Tab. \ref{tab:systemic_velocity}.  By folding the
major axis slit (which reaches furthest out in $R$) about the origin, after 
subtracting out $V_{\it sys}$, we obtained the stellar streaming velocities, 
$V_*(R)$, shown in Fig. \ref{fig:circular_velocity}.

\begin{table}
\caption{Heliocentric systemic velocity based on optical measurements.} 
\begin{tabular}{ll} 
\hline 
\multicolumn{1}{c}{$V_{\it sys}$} & Source      \\
\multicolumn{1}{c}{(\kms)}        &      \\
\hline 
$734 \pm 41$ & Mayall \& de Vaucouleurs (1962) \\
$617 \pm 19$ & Tonry \& Davies (1981)\\
$597 \pm  9$ & Schechter (1983)\\
$608 \pm  9$ & Schechter (1983)\\
$615 \pm 20$ & Dressler \& Sandage (1983)\\
$620 \pm 50$ & Capaccioli \etal\ (1986)\\
$592 \pm 15$ & Simien \& Prugniel (1997)\\
\hline
$606 \pm 5$ & Weighted average of literature values \\
$613 \pm 3$ & This work \\
\hline
\label{tab:systemic_velocity}
\end{tabular}
\end{table}

We tested $V_*(R)$ by rotating it
into the orientations of other slit observations available in the
literature.  The slits we compared with are those at PA $=87\degrees$
(Simien \& Prugniel 1997; Neistein \etal\ 1999), PA $=90\degrees$
(Bower \etal\ 2001) and PA $=177\degrees$ (Simien \& Prugniel 1997).
Excluding the inner $2\arcsec$, to avoid the effects of different
seeing and/or slit widths, we find $\tilde{\chi}^2 \ltsim 1$ in all
cases except for the PA $=87\degrees$ slit of Neistein \etal\ (1999),
for which we obtained $\tilde{\chi}^2 = 2.5$ (for 27 points).  Since
the same PA is also covered by a slit from Simien \& Prugniel (1997),
which we match well ($\tilde{\chi}^2 = 0.8$ for 36 points), we
conclude that the stellar streaming velocity profile we obtained is
reasonable and in good agreement with previous data.

\begin{figure}
 \leavevmode{\psfig{figure=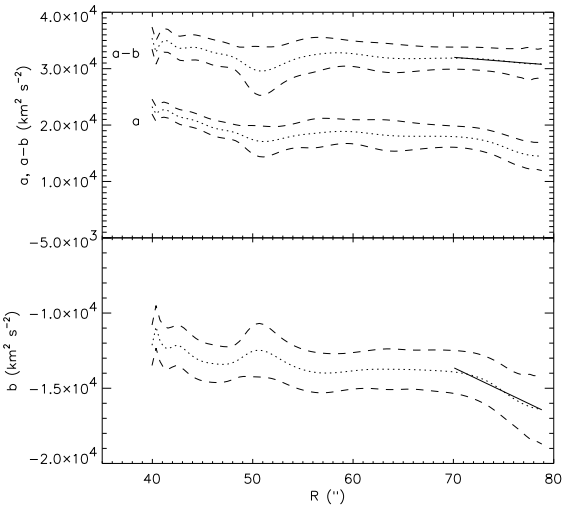,height=7.5truecm,width=7.5truecm}}
\caption[]{The average value, $a$, (top panel) and the $\cos 2\phi$ 
  component, $b$, (bottom panel) of the squared observed velocity
  dispersions.  The top panel also indicates $a-b$.  We fit straight
  lines to $a-b$ and $b$ at $R \geq 70\arcsec$, as indicated.} 
\label{fig:ab_plot}
\end{figure}

We then corrected for the asymmetric drift using the velocity dispersion 
data, to recover the true circular velocity, $V_c(R)$.  Following Eqn. 4-33 
in Binney \& Tremaine (1987), and assuming 
$\partial(\overline{v_R v_z})/\partial z \simeq 0$, we can write the 
asymmetric drift equation as:
\begin{equation}
V_c^2 - V_*^2 = -\sigma_R^2 \left[ 
{\frac{\partial \ln \rho}{\partial \ln R}} +
{\frac{\partial \ln \sigma_R^2}{\partial \ln R}} + \left( 1 - 
{\frac{\sigma_\phi^2}{\sigma_R^2}} \right) \right].
\end{equation}
The observed velocity dispersion, at a point $(R,\phi)$ in the galaxy frame, 
is given by:
\begin{eqnarray}
\sigma^2_{los} & = & \frac{1}{2} \sin^2i\left[(\sigma^2_R + \sigma^2_\phi +
2 \sigma^2_z \cot^2i) - (\sigma^2_R - \sigma^2_\phi) \cos 2\phi 
\right] \nonumber \\
& \equiv & a(R,\phi) + b(R,\phi) \cos 2\phi.
\end{eqnarray}
where $\phi = 0$ is taken to be the disc's major axis.  When the disc
is axisymmetric, $a$ and $b$ can only depend on $R$, in which case 
$a$ and $b$ are simply the first 2 even moments in a Fourier expansion
of the line-of-sight velocity dispersion.  For a disc with a flat rotation
curve, $\sigma^2_\phi/\sigma^2_R = 1/2$, which then allows us to
obtain $\sigma^2_R$ from $b$.  However, the profile of $V_*$ we found
for NGC 1023 is not obviously flat.  (Furthermore, $b$ is more
sensitive to noisy data than is $a$.)  In the general case, the
asymmetric drift equation can be re-written in terms of both $a$ and
$b$, as follows:
\begin{eqnarray}
V_c^2 - V_*^2 & = &
\frac{R}{\sin^2 i (1 + \alpha^2 \cot^2 i)} \left[
\frac{a-b}{R_{\it d}} - \frac{\partial}{\partial R}(a-b)
\right]
\nonumber \\
& + & \frac{2b}{\sin^2 i},
\end{eqnarray}
where we have assumed that the disc density is exponential with
scale-length $R_{\it d}$ and $\alpha \equiv \sigma_z/\sigma_R$.  Using
the velocity dispersion data from our 3 slits and the near minor-axis
data of Simien \& Prugniel (1997), we measured $a-b$ and $b$.  We
fitted splines to interpolate between data points, and estimate errors
from Monte Carlo simulation with Gaussian noise added from the error
estimate on each point.  The resulting profiles of $a$ and $b$ are
shown in Fig. \ref{fig:ab_plot}. Our procedure for measuring $V_*(R)$,
$a$ and $b$ is valid only in the axisymmetric region outside the bar.
Therefore we only apply the asymmetric drift correction to $R >
70\arcsec$; for this region, we fit straight lines to $a-b =
-141\,(\pm 61)\,R + 41884\,(\pm 562)$ km$^2$ s$^{-2}$ and to $b =
-319\,(\pm 39) R + 8706\,(\pm 350)$ km$^2$ s$^{-2}$.  With these
values, we then correct the rotation curve, assuming $0.0 \leq \alpha
\leq 1.0$.  (This range of $\alpha$ is a reasonable one: in the Solar 
neighborhood, Dehnen \& Binney (1998) found $\alpha \simeq 
0.54^{+0.02}_{-0.05}$, while in NGC 488, Gerssen \etal\ (1997) measured 
$\alpha = 0.70 \pm 0.19$.)  The resulting rotation curve is more or less 
flat, with $259 \ltsim V_{c,flat} \ltsim 278$ \kms.  Adding in quadrature a
velocity uncertainty of 30 \kms, we obtain $V_{\it c,flat} = 270 \pm
31$ \kms.  (For comparison, Neistein \etal\ 1999 found $V_c = 250 \pm
17$ \kms; we have checked that, using a method similar to their Eqn.
8, gives similar values for $V_{c,flat}$ to those reported above.)  If
we extrapolate this value to smaller radii (which, at worst, over-estimates 
$R_c$ and, correspondingly, $\vpd$, unless the rotation curve is
dropping between $40\arcsec \ltsim R \leq 70\arcsec$, which seems unlikely
from Fig. \ref{fig:circular_velocity}) we find 
$R_c = V_{\it c,flat}/\om = {53 \pm 6 ^{+29}_{-14}} \arcsec$, where the first
error is due to the uncertainties in $V_{c,flat}$ and the second error
reflects uncertainties in $\om$.  We obtain, therefore, 
$\vpd = 0.77 \pm 0.10 ^{+0.42}_{-0.20}$ (where the first error now also 
includes the uncertainty in $\len$).  For our ``slowest-bar'' estimate of 
the slope, we find $\vpd = 0.91 \pm 0.12 ^{+0.49}_{-0.23}$.  We have found,
therefore, that the bar in NGC 1023 is consistent with being a fast bar.

\section{Discussion and Conclusions}
\label{sec:conclusions}

We have found that the bar in NGC 1023 is fast, as are all bars which 
have been measured to date.  Debattista \& Sellwood (1998, 2000) showed that 
fast bars can persist only if the disc is maximal.  Following Ostriker \& 
Peebles (1973), it is sometimes thought that the unbarred (SA) galaxies are 
stabilized by massive DM halos.  However, massive DM halos are not
necessary for stabilizing discs; a rapidly rising rotation curve in the 
inner disc, such as when a massive bulge is present, is also able to inhibit 
bar formation (Toomre 1981; Sellwood \& Evans 2001).  Debattista \& Sellwood 
(1998) argued that unbarred 
HSB galaxies must also be maximal for, if HSB disc galaxies form a continuum 
of DM halo masses spanning massive DM halo-stabilised SA galaxies to maximal 
SB galaxies, then slow bars must also be found in the intermediate range of
halo masses.  If we seek to avoid intermediate halo masses and slow bars by 
postulating (for whatever reason) a bimodal DM halo mass distribution for HSB 
galaxies, then we are left with the possibility that tidal interactions can
still form bars, which would be slow (Noguchi 1987; Salo 1991; Miwa \& 
Noguchi 1998).
Thus Debattista \& Sellwood (2000) concluded that the absence of 
slow bars requires that all HSB disc galaxies are maximal.  However, it is
possible that no such slow bars have been found because of an observational 
bias against SB systems with evidence of tidal interactions.

We have chosen to study NGC 1023, in part, because it shows signs of a weak 
interaction in its past, without being at present significantly perturbed.  
The fast bar we found indicates that NGC 1023 has a maximal disc.  If SA 
galaxies are stabilized by massive halos, we should find slow bars in that
fraction of SB galaxies in which the bar formed through the interaction.
While it is not possible to reach a general conclusion on the DM content of 
SA galaxies based on our measurement for a single galaxy, a large enough 
sample of similar SB galaxies with mild interactions in the past will be 
able to address this question.

\bigskip
\noindent
{\bf Acknowledgements.} 

\noindent
V.P.D. and J.A.L.A. acknowledge support by the Schweizerischer
Nationalfonds through grant 20-56888.99.  V.P.D. wishes to thank the
Dipartimento di Astronomia dell'Universit\`a di Padova for hospitality
while preparing for the observations.  E.M.C.  acknowledges the
Astronomisches Institut der Universit\"at Basel for the hospitality
while this paper was in progress.  We wish to thank the staff of the
JKT telescope, particularly to the support astronomer J. C. Vega-Beltr\'an 
and the staff of the TNG.  We are indebted to R. Bender and R. Saglia for 
providing us with the FCQ package which we used for measuring the stellar 
kinematics.  We thank the anonymous referee for suggestions that helped
improve this paper.  This research has made use of the Lyon-Meudon 
Extragalactic Database (LEDA) and of the NASA/IPAC Extragalactic 
Database (NED).


\bigskip
\noindent

\begin{thebibliography}{}

\bibitem{} Aguerri J.\ A.\ L., Mu\~noz-Tu\~n\'on C., Varela A.\ M., 
Prieto M.\ 2000, A\&A, 361, 841.

\bibitem{} Aguerri J.\ A.\ L., Hunter J.\ H., Prieto M., Varela A.\ M., 
Gottesman S.\ T., Mu\~noz-Tu\~n\'on C.\ 2001, A\&A, 373, 786.
  
\bibitem{} Arp H.\ 1996, Atlas of Peculiar Galaxies.  California
  Institute of Technology, Pasadena

\bibitem{} Athanassoula E.\ 1992, MNRAS, 259, 345
  
\bibitem{} Athanassoula E.\ 1996, in IAU Colloq.\ 157, Barred Galaxies, ed.\ 
R.\ Buta, D.\ A.\ Crocker, \& B.\ G.\ Elmegreen (PASP Conf series 91), 309

\bibitem{} Barbon R., Capaccioli M.\ 1975, A\&A, 42, 103

\bibitem{} Bender R.\ 1990, A\&A,  229, 441 
  
\bibitem{} Bender R., Saglia R.\ P., Gerhard O.\ E.\ 1994, MNRAS, 269,
  785 
  
\bibitem{} Binney J., Tremaine S.\ 1987, Galactic Dynamics, Princeton
  University Press, Princeton
  
\bibitem{} Bower G.\ A., Green R.\ F., Bender R., \etal\ 2001, ApJ, 550,
  75

\bibitem{} Capaccioli M., Lorenz H., Afanasjev V.\ L.\ 1986, A\&A, 169,
  54

\bibitem{} Cardelli J.\ A., Clayton G.\ C., Mathis J.\ S.\ 1989, ApJ, 345, 245

\bibitem{} Contopoulos G.\ 1980, A\&A, 81, 198

\bibitem{} Debattista V.\ P.\ 1998, Ph.D. Thesis, Rutgers University

\bibitem{} Debattista V.\ P. 2002, {\it in preparation}

\bibitem{} Debattista V.\ P., Sellwood J.\ A.\ 1998, ApJ, 493, L5

\bibitem{} Debattista V.\ P., Sellwood J.\ A.\ 2000, ApJ, 543, 704

\bibitem{} Debattista V.\ P., Williams T.\ B.\ 2001, in Galaxy Disks and 
Disk Galaxies, ed. J. G. Funes \& E. M. Corsini, ASP Conf. Ser. 230, ASP, 
San Francisco, 553

\bibitem{} Debattista V.\ P., Williams T.\ B.\ 2002, {\it in preparation}
  
\bibitem{} Dehnen W., \& Binney J.\ J.\ 1998, MNRAS, 298, 387

\bibitem{} de Vaucouleurs G., de Vaucouleurs A., Corwin H.\ G., Buta R.\
  J., Paturel G., Fouqu\`e P.\ 1991, Third Reference Catalogue of Bright
  Galaxies.  Springer-Verlag, New York (RC3)

\bibitem{} Elmegreen B. 1996, in IAU Colloq.\ {\bf 157}, Barred Galaxies, 
ed. R. Buta, D. A. Crocker \& B. G. Elmegreen (San Francisco: ASP Conf series 
91), 197

\bibitem{} Elmegreen B.\ G., Elmegreen D.\ M.\ 1990, ApJ, 355, 52

\bibitem{} England M.\ N., Gottesman S.\ T., Hunter J.\ H.\ 1990, ApJ, 348, 456
  
\bibitem{} Eskridge P.\ B., Frogel J.\ A., Pogge R.\ W., \etal\ 2000, AJ,
  119, 536

\bibitem{} Faber S.\ M., Tremaine S., Ajhar E.\ A., \etal\ 1997, AJ,
  114, 1771
  
\bibitem{} Garcia A.\ M.\ 1993, A\&AS, 100, 47

\bibitem{} Gerhard O.\ E.\ 1993, MNRAS,  265, 213 

\bibitem{} Gerssen J., Kuijken K., Merrifield M.\ R.\ 1997, MNRAS, 288, 618

\bibitem{} Gerssen J., Kuijken K., Merrifield M.\ R.\ 1999, MNRAS, 306,
  926

\bibitem{} Hernquist, L., Weinberg, M.\ D.\ 1992, ApJ, 400, 80

\bibitem{} Hunter J.\ H., Ball R., Huntley J.\ M., England M.\ N., 
Gottesman S.\ T.\ 1989, ApJ, 324, 721

\bibitem{} Kent, S.\ M.\ 1987, AJ, 93, 1062

\bibitem{} King D.\ L.\ 1985, La Palma Technical Note, N 31

\bibitem{} Kormendy J.\ 1985, ApJ,  292, L9 

\bibitem{} Knapen J.\ H.\ 1999, in The Evolution of Galaxies on Cosmological
Timescales, ed.\ J. E. Beckman, \& T. J. Mahoney (San Francisco: ASP 
Conference Series 187), 72

\bibitem{} Knapen J.\ H., Shlosman I., Peletier R.\ F.\ 2000, ApJ, 529, 93

\bibitem{} Laine S.\ 1996, Ph. D. Thesis, Florida University.
  
\bibitem{} Lauer T.\ R., Ajhar E.\ A., Byun Y.-I., \etal\ 1995, AJ, 110,
  2622

\bibitem{} Lindblad P.\ A.\ B., Kristen H.\ 1996, A\&A, 313, 733
  
\bibitem{} Lindblad P.\ A.\ B., Lindblad P.\ O., Athanassoula E.\ 1996,
  A\&A, 313, 65

\bibitem{} Little, B., Carlberg, R.\ G.\ 1991, MNRAS, 250, 161
 
\bibitem{} Merrifield M.\ R., Kuijken K.\ 1995, MNRAS, 274, 933

\bibitem{} Miwa T., Noguchi M.\ 1998, ApJ, 499, 149
  
\bibitem{} M\"ollenhoff C., Heidt J.\ 2001, A\&A, 368, 16

\bibitem{} Navarro J.\ F., Frenk C.\ S., White S.\ D.\ M.\ 1997, ApJ, 490, 493

\bibitem{} Neistein E., Maoz D., Rix H.-W., Tonry J.\ L.\ 1999, AJ, 117,
  2666
  
\bibitem{} Nilson P.\ 1973, Uppsala General Catalogue of Galaxies.
  Royal Society of Sciencies of Uppsala, Uppsala

\bibitem{} Noguchi M.\ 1987, MNRAS, 228, 635

\bibitem{} Osterbrock D.\ E., Fulbright J.\ P., Martel, A.\ R., Keane
  M.\ J., Trager S.\ C., Basri, G.\ 1996, PASP, 108, 277

\bibitem{} Ostriker J.\ P., Peebles P.\ J.\ E.\ 1973, ApJ, 186, 467

\bibitem{} Ohta K., Masaru H., Wakamatsu K.\ 1990, ApJ, 357, 71
  
\bibitem{} Pogge R.\ W., Eskridge P.\ B.\ 1993, AJ, 106, 1405
  
\bibitem{} Salo H.\ 1991, A\&A, 243, 118

\bibitem{} S\'anchez-Portal M., D\'\i az \'Angeles I., Terlevich R.,
  Terlevich E., \'Alvarez \'Alvarez M., Aretxaga I.\ 2000, MNRAS, 312,
  2

\bibitem{} Sancisi R., van Woerden H., Davies R.\ D., Hart L.\ 1984,
  MNRAS, 210, 497

\bibitem{} Sandage A., Tammann G.\ A.\ 1981, A revised Shapley-Ames
  Catalog of Bright Galaxies, Carnegie Institution of Washington, 
  Washington D.C.

\bibitem{} Schelgel D.\ J., Finkbeiner D.\ P., Davis M.\ 1998, ApJ, 500, 525

\bibitem{} Sellwood J.\ A.\ 1980, A\&A, 89, 296

\bibitem{} Sellwood J.\ A., Wilkinson A\ 1993, Rep. Prog. Phys., 56, 173

\bibitem{} Sellwood J.\ A., Evans N.\ W.\ 2001, ApJ, 546, 176

\bibitem{} Sil'chenko O.\ K.\ 1999, AJ, 117, 2725
  
\bibitem{} Simien F., Prugniel P.\ 1997, A\&AS, 126, 519
  
\bibitem{} Sofue Y., Wakamatsu K., Taniguchi Y., Nakai N.\ 1993, PASJ,
  45, 43

\bibitem{} Taniguchi Y., Murayama T., Nakai N., Suzuki M., Kameya O.\
  1994, AJ, 108, 468

\bibitem{} Toomre A.\ 1981, in The Structure and Evolution of Normal 
Galaxies, ed.\ S. M. Fall, \& D. Lynden-Bell (Cambridge: Cambridge University 
Press), 111

\bibitem{} Tremaine S., Weinberg M.\ D.\ 1984, ApJ, 282, L5

\bibitem{} Tremaine S., Ostriker J.\ P.\ 1999, MNRAS, 306, 662

\bibitem{} van Albada T.\ S., Sanders R.\ H.\ 1982, MNRAS, 201, 303
  
\bibitem{} van Albada T.\ S., Sancisi R.\ 1986, Phil.\ Trans.\ R. Soc.\ 
  Lond.\ A, 320, 447

\bibitem{} van der Marel R.\ P., Franx M.\ 1993, ApJ,  407, 525 

\bibitem{} Weinberg M.\ D.\ 1985, MNRAS, 213, 451
  
\bibitem{} Weiner B.\ J., Sellwood J.\ A., Williams T.\ B.\ 2001, ApJ,
  546, 931

\end{thebibliography}
\end{document}